\documentclass[aps,twocolumn,groupedaddress,showpacs,showkeys]{revtex4-1}

\usepackage{amssymb,amsmath} 
\usepackage[bb=boondox]{mathalpha}
\usepackage{newtxtext,newtxmath}
\usepackage[linktoc=page]{hyperref}
\usepackage{color}
\usepackage{graphicx}
\usepackage{units}
\usepackage{verbatim}
\usepackage{bbold} 

\def\Omg{\Omega}
\def\omg{\omega}

\def\kcoord{k}
\def\vcoord{v}
\def\tcoord{t}
\def\xcoord{x}
\def\wcoord{w}
\def\zcoord{z}

\def\Rcoord{R}
\def\Vcoord{V}
\def\Zcoord{Z}

\def\ahat{\mathbf {\hat a}}
\def\bhat{\mathbf {\hat b}}
\def\chat{\mathbf {\hat c}}

\def\Vk{\mathbf k}

\def\Vp{\mathbf p}
\def\Vu{\mathbf u}
\def\Vw{\mathbf w}
\def\Vx{\mathbf x}

\def\VA{\mathbf A}
\def\VB{\mathbf B}
\def\VC{\mathbf C}
\def\VE{\mathbf E}
\def\VJ{\mathbf J}

\def\VR{\mathbf r}

\def\Vv{\mathbf v}
\def\VV{\mathbf V}

\def\Vvel{\mathbf v}

\def\VF{\mathbf{F}}

\def\Vp{{\bf p}}
\def\Vd{{\bf d}}

\def\VC{{\bf C}}
\def\VR{{\bf R}}
\def\VS{{\bf S}}

\def\CJ{{\mathcal J}}
\def\CK{{\mathcal K}}

\def\CO{{\mathcal O}}
\def\CT{\mathcal{T}}

\def\SG{\mathsf{G}}
\def\SJ{\mathsf{J}}
\def\ST{\mathsf{T}}

\def\STzero{\ST}

\def\Vnabla{\boldsymbol \nabla}
\def\Vkappa{\boldsymbol{\kappa}}

\def\unitTensor{{\bf 1}}
\def\porder{p}

\def\nindex{n}
\def\mindex{m}

\def\EM{EM}
\def\GC{{\rm GC}}

\def\cyc{\mathrm{c}}

\def\drift{\mathrm{d}}

\def\tot{\mathrm{total}}

\def\drift{{\rm drift}}
\def\pol{{\rm pol}}

\def\free{\mathrm{drift}}
\def\bound{\mathrm{pol}}
\def\bound{\mathrm{bound}}
\def\free{\mathrm{free}}

\def\reference{reference }
\def\eff{\rm esc}

\def\Omgc{\Omega_c}
\def\Omgc{\Omega}

\def\gyro{{\rho}}

\def\Vgyro{\boldsymbol \gyro}

\def\Ham{\Hamiltonian}
  
\def\Pphi{P_\phitor}

\def\Jbar{{\bar\Jcyc}}
\def\VRbar{{\bar\VR}}

\def\chargedensity{\varrho}
\def\potential{\phi}

\def\Afield{A}

\def\Bfield{B}
\def\VBfield{\mathbf \Bfield}
\def\Efield{E}
\def\VEfield{\mathbf \Efield}
\def\Jfield{J}

\def\Apar{\Afield_\|}

\def\Bpar{\Bfield_\|}

\def\Epar{\Efield_\|}
\def\Eperp{\Efield_\perp}
\def\VEperp{\mathbf{\Eperp}}

\def\VD{\mathbf{D}}
\def\VH{\mathbf{H}}

\def\Vpol{\boldsymbol{\pi}}
\def\VL{\varmathbb{L}}
\def\VMag{\varmathbb{M}}
\def\VPol{\varmathbb{P}}
\def\VQuad{\varmathbb{Q}}

\def\dipole{p}
\def\Vd{\mathbf{d}}
\def\Vdipole{\boldsymbol{\mathsf{p}}}
\def\Vpol{\boldsymbol{\pi}}
\def\Vquad{\boldsymbol{\mathsf{q}}}
\def\VL{\boldsymbol{\mathsf{L}}}
\def\VMag{\boldsymbol{\mathsf{M}}}
\def\VPol{\boldsymbol{\mathsf{P}}}
\def\VQuad{\boldsymbol{\mathsf{Q}}}
\def\Squad{q}

\def\kpar{k_\|}
\def\kperp{k_\perp}

\def\Jac{\mathcal{J}}

\def\kperp{k_\perp}
\def\Vkperp{\Vk_\perp}

\def\pol{{\rm pol}}

\def\phitor{\varphi} 
\def\phitor{\varphi}

\def\Vpar{u_\|}

\def\mass{m}
\def\charge{q}

\def\pdf{f}

\def\PDF{F}
\def\gfun{g}
\def\Gfun{G}

\def\density{n}
\def\press{p}
\def\Temp{T}
\def\Tpar{T_\|}
\def\Tperp{T_\perp}

\def\Density{N}
\def\Press{P}

\def\pressTensor{{\bf \press}}
\def\PressTensor{{\bf \Press}}

\def\diamag{ * }
\def\kin{K}
\def\kinetic{thermodynamic }

\def\Upar{U_\|} 
\def\vel{v}
\def\vpar{\vel_\|}
\def\vperp{\vel_\perp}
\def\Vvel{v}

\def\Vvperp{\Vvel_\perp}

\def\Vel{V}
\def\Vpar{\Vel_\|}

\def\Action{S}
\def\Lagrangian{L}
 
\def\Ham{H}

\def\Action{S}
\def\Jcoord{J} 
\def\gyrophase{\theta}
\def\Jcyc{{\Jcoord_\cyc}}

\def\GKcoord{{\bar Z}}

\def\Jcyc{ \Jcoord }

\def\Rcoord{X}
\def\VRcoord{\mathbf{\Rcoord}}
\def\VR{\mathbf{\Rcoord}}
\def\Rbar{{\bar \Rcoord}}

\def\CKavg{{\overline{ \CK} } }
\def\CKvary{{\widetilde{ \CK}  }} 
\def\Source{S}
\def\ColOp{C}
\def\Cop{C}
\def\CollOp{C}

\def\epsperp{\epsilon_\perp}
\def\epspar{\epsilon_\|}
\def\epsnonlin{\delta}

\def\zeroflr{0} 
\def\firstflr{1}
\def\secondflr{2} 

\def\eq{{\rm eq}} 
\def\pert{\delta}

\def\densitystart{\density_\zeroflr}
\def\pressstart{\press_{\zeroflr \perp}}
\def\pdfstart{\pdf_\zeroflr}
\def\densitystartpert{\pert\densitystart}
\def\pdfstartpert{\pert\pdfstart}
\def\Presspert{\pert\Press}
\def\PDFbar{{\bar \PDF}}

\def\Bparpert{\pert \Bpar} 
\def\Aparpert{\pert \Apar} 
\def\chipert{\pert\chi}

\def\pdfpert{\pert \pdf}
\def\PDFpert{\pert \PDF}
\def\phipert{\pert \phi}
\def\densitypert{\pert\density}
\def\Densitypert{\pert \Density}
\def\Temppert{\pert \Temp}
\def\presspert{\pert \press}

 \def\Epert{\pert E}
 
 \def\VVpert{\pert\VV}

\newcommand\abs[1]{\left|#1\right|}

\newcommand{\avg}[1]{{ \left< #1 \right>}}
\renewcommand{\vary}[1]{{ \left\{ #1 \right\}}}

\def\radial{ \zcoord }
\def\cyc{\Zcoord}

\newcommand{\avgVel}[1]{{ \avg{#1}_\radial}}
\newcommand{\varyVel}[1]{{ \vary{#1}_\radial}}

\newcommand{\avgCyc}[1]{{ \avg{#1}_\cyc }}
\newcommand{\varyCyc}[1]{{ \vary{#1}_\cyc}}

\newcommand{\avgCycBar}[1]{{ \avg{\overline{#1}}_{\bar \cyc} }}
\newcommand{\varyCycBar}[1]{{ \vary{\overline{#1}}_{\bar \cyc}}}

\def\PDFvary{\varyCyc{\PDF}}
\def\Svary{\varyCyc{\Source}}

\begin{document}


\title{ A Tale of Two Polarization Paradoxes I: \\
the Diamagnetic Polarization Paradox }


\author{Ilon Joseph}
\email[]{joseph5@llnl.gov}
\homepage[]{https://people.llnl.gov/joseph5}

\affiliation{Lawrence Livermore National Laboratory}


\date{\today}

\begin{abstract}
An accurate calculation of the total polarization charge density in a plasma is essential for a self-consistent determination of the electric field. 
Yet, for a magnetized plasma, there are two different  polarization paradoxes that are finally resolved in this work.
The ``diamagnetic polarization paradox'' refers to the fact that there is a paradoxical factor of 1/2 difference between the pressure-driven ``diamagnetic polarization'' density  calculated using the real space drift theory versus the action-angle space guiding center and gyrokinetic  theory that has not been explained before. 
In this work, it is shown that the results of both approaches can be made  consistent with one another.
Half of the diamagnetic polarization is due to the transformation from the guiding center density to the real space density. 
The other half is due to the fact that, within the drift kinetic ordering assumptions, the guiding center density should be expressed as the gyroaverage of the density in the limit of vanishing Larmor radius.
Expressions for the diamagnetic polarization density are given that are accurate to first order in amplitude and all orders in gyroradius within gyrokinetic theory for a constant magnetic field.
Applications to anisotropic Maxwell-Boltzmann particle distribution functions 
are presented.
Because the total energy and toroidal momentum are local invariants, they
do not generate net polarization effects: the electric and thermodynamic polarizations must precisely cancel. In contrast, anisotropic dependance on the magnetic moment
generates a net polarization proportional to the temperature anisotropy.
\end{abstract}

\maketitle



\tableofcontents

\def\mag{\mathrm{mag}}  

\section{Introduction} 
\subsection{A Tale of Two Paradoxes}
The polarization of charged particles in a strong magnetic field plays a central role in the adiabatic theory of magnetized plasmas.  
The dependence of the polarization charge density on the electric field allows one to solve for the electric potential through the so-called ``vorticity'' equation \cite{Zeiler97pop, Xu00nf, Xu00pop, Simakov03pop, Simakov04pop, Parra09ppcf_vorticity} or the ``gyro-Poisson'' equation \cite{LeeWW83pof, Dubin83pof, Brizard07rmp, Brizard08cnsns,  Parra10ppcf_toroidal, Parra11ppcf, Brizard13pop, Brizard24jpp}. 
The \emph{electric polarization} is always a key part of the determination of the electric field.
But, in fact, both an electric field and a pressure gradient generate a polarization density and both effects are needed to determine the electric field self-consistently. 
The effect of the electric field is due to the single-particle electric drift, while the effect of the pressure gradient is due to the collective fluid diamagnetic drift.   
The diamagnetic drift is not a single particle drift; rather, it is generated by the magnetization current that arises from the finite Larmor radius (FLR) of the particle orbits.
Since the pressure-gradient driven polarization is due to the polarization current induced by the diamagnetic drift,  it will be referred to as the \emph{diamagnetic polarization}.
In principle, other thermodynamic forces, such as temperature and velocity gradients, also contribute to the polarization.
Hence, we distinguish the electric polarization, which is defined as the kinetic polarization due to the electromagnetic fields, from the  \emph{ \kinetic polarization}, which is defined as the kinetic polarization due to thermodynamic forces.

There is a paradoxical factor of 1/2 difference between the magnitude of the  diamagnetic polarization derived using the standard real phase space drift-reduced approach versus the action-angle guiding center and gyrokinetic approaches \cite{Bolton16pvt}.  
The real phase space approach uses standard position and velocity space coordinates, $\zcoord=\{\xcoord,\vcoord\}$, and is used to derive the results for the ``drift-reduced kinetic'' (DK)  equation \cite{Hazeltine1973pp, Hazeltine92book, Helander02book}
and the  ``drift-reduced MHD'' and ``drift-reduced Braginskii'' fluid equations (collectively denoted DF)
\cite{Spitzer1952apj, Hazeltine92book, Simakov03pop, Simakov04pop}.  
In contrast, the transformation to adiabatic action-angle coordinates, $\Zcoord=\{\Rcoord,\Vpar,\Jcyc, \gyrophase\}$, is used to derive guiding center (GC) \cite{Littlejohn81pof,Littlejohn83jpp,Tronko15pop,Brizard16arxiv,Brizard24jpp}, gyrokinetic (GK) \cite{LeeWW83pof, Dubin83pof} and gyrofluid (GF) \cite{Brizard92pfb, Dorland93pfb} results.   
Although a number of authors have derived linear and nonlinear  expressions for the  polarization density \cite{Brizard07rmp, Brizard08cnsns, Parra10ppcf_toroidal, Parra11ppcf, Brizard13pop, Tronko15pop, Brizard16arxiv, Sugama22pop, Brizard24jpp}, there does not seem to be a clear explanation of this discrepancy in the literature. Since the difference is in the linear polarization at second order in the ratio of gyroradius to wavelength,  the issue is not due to nonlinear terms nor due to  finite Larmor radius (FLR) effects that are higher than second order.

In this work, we refer to the fact that the diamagnetic polarization density  appears to be different in the real space DK/DF and the adiabatic GC/GK/GF  pictures as the \emph{diamagnetic polarization paradox}.
We prove that, when the same assumptions are used, the paradox is only apparent and is not an actual discrepancy.
\emph{\bf The key difference lies in which \reference distribution the density is being compared to}.
The paradox is resolved by retaining all second order terms  in the translation between the GC/GK/GF and DK/DF pictures, including the \reference PDFs.

In fact, there is a close connection to the \emph{Spitzer magnetization paradox} \cite{SpitzerBook}, the fact that the single-particle drifts do not agree with the collective fluid particle drifts until the magnetization current, due to the finite Larmor radius of the motion of particles around the guiding center, is also taken into account.
The polarization density is determined by the polarization current, which, in turn, is determined by the drift due to inertial acceleration.
Both the single particle drift and the magnetization current are required to determine the full polarization current and, hence, the full polarization density.
Thus, there is also a \emph{Spitzer polarization paradox}: the single-particle polarization drifts and density do not agree with the collective fluid particle polarization drifts and density until the inertial response to the magnetization current is also taken into account.
Part II of this series \cite{Joseph2025dkgc-II} will resolve the Spitzer polarization paradox.

\begin{figure}
\includegraphics[width=2in]{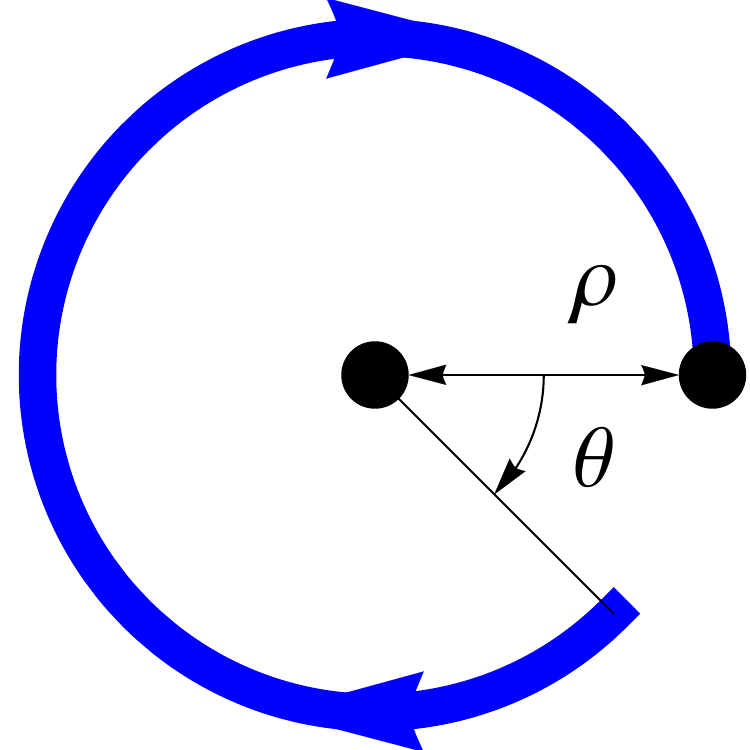}
\caption{The finite size of the gyroradius, $\gyro=\vperp/\Omgc$, is responsible for diamagnetic polarization effects.}
\label{fig:Larmor_orbit}
\end{figure}

The diamagnetic polarization paradox is a separate and distinct issue that will be resolved in Part I of this series. 
The  polarization density is usually derived in a form that is accurate to second order in FLR effects and first order in amplitude.  This approximation requires the characteristic wavelengths to be much longer than the  gyroradius, $\gyro=\vperp/\Omgc$, and the characteristic frequencies to be much smaller than the gyrofrequency, $\Omgc=\charge\Bfield/\mass$, for a particle with charge, $\charge$, and mass, $\mass$, in a magnetic field, $\VB$. 
The polarization density can be expressed as
\begin{align}\label{eq:pol_drift}
{\rm DK/DF: } &&  \density_\pol (\xcoord)  &=    \Vnabla\cdot   \left[ (\Vnabla \cdot\pressTensor)_\perp- \charge\density \VE_\perp \right] /\mass\Omgc^2 \\
{\rm GC/GK/GF:} &&     \Density_\pol(\xcoord) &=  \left. \Vnabla  \cdot   \left[( \tfrac{1}{2} \Vnabla\cdot \PressTensor )_\perp-\charge\Density \VE_\perp  \right] /\mass\Omgc^2\right|_{\xcoord}
\label{eq:pol_gyro}
.
\end{align}
 The first term on the right hand side of both equations is the diamagnetic polarization, while the second term is the electric polarization, where $\VE_\perp$, is the electric field perpendicular to $\VB$.  In Eq.~\ref{eq:pol_drift}, $\density$ is the particle density and $\pressTensor$ is the pressure tensor in real space coordinates, $\xcoord$.    
 In Eq.~\ref{eq:pol_gyro}, $\Density$ is the guiding center  density and   $\PressTensor$ is the guiding center pressure tensor,  defined as a function of guiding center coordinates, $\Rcoord$, (defined in Appendix~\ref{sec:gc_coordinates}) but evaluated at the real space coordinate position, $\xcoord$. 
 Although the pressure gradient term in Eq.~\ref{eq:pol_gyro} differs by a factor of 1/2 from that in Eq.~\ref{eq:pol_drift}, both results are actually consistent with one another when the same set of assumptions is used.
 The differences are clearly due to the displacement between the guiding center position and the true particle position illustrated in Fig.~\ref{fig:Larmor_orbit}.

To this author's knowledge, the first reference to Eq.~\ref{eq:pol_drift} is L. Spitzer, Jr.'s discussion of an ideal plasma  \cite{Spitzer1952apj}.
We review the derivation of this result through an analysis of the drift fluid equations in Sec.~\ref{sec:drift-fluid}.
Drift kinetic theory traces its origins to the seminal studies of drift instabilities \cite{Hastie1967ap, Rutherford1967pof, Rutherford1968pof}, collisional transport \cite{Rutherford1970pof, Frieman1970pof}, and neoclassical transport \cite{Rosenbluth72pof, Hazeltine1973pof, Hinton1976rmp}.
It was put on more rigorous footing by R. D. Hazeltine \cite{Hazeltine1973pp, Hazeltine1978pp} and, today, there are excellent expositions in \cite{Hazeltine92book, Helander02book}.
However, the theory is often only presented to first order in $\kperp\gyro$, which is why the second order polarization terms are not as widely appreciated.

Modern derivations of drift kinetic theory \cite{Hazeltine92book, Helander02book} use the real space approach, but then go on to explain that the GC/GK approach of transforming to adiabatic coordinates is a more rigorous interpretation of the procedure. 
As, Hastie, et al,  \cite{Hastie1967ap} explained,if one neglects collisions and sources, kinetic theory can be used to find corrections to adiabatic invariants.
However, in traditional DK/DF theory, the results are calculated in real space rather than in adiabatic coordinates and this causes differences between the two forms of the perturbation series.
 
The importance of the second order GK result in the derivation of the gyro-Poisson equation was first described by W. W. Lee \cite{LeeWW83pof} and was first rigorously extended to include higher order FLR effects and nonlinear terms by D. H. E. Dubin, et al., \cite{Dubin83pof}.    Equation~\ref{eq:pol_gyro} was originally derived for the equilibrium density, but it is also true for perturbations.
This result was first derived from the gyrofluid perspective by A. J. Brizard~\cite{Brizard92pfb}. Modern GC and GK results, which have been extended to include nonlinear terms \cite{Brizard07rmp, Brizard08cnsns, Parra09ppcf_gkequiv, Parra11ppcf, Brizard13pop, Tronko15pop, Brizard16arxiv, Sugama22pop, Brizard24jpp}, still agree with  Eq.~\ref{eq:pol_gyro} at this order. However, we note that Refs.~\onlinecite{LeeWW17pop,LeeWW18pop,LeeWW19pop} have attempted to  use the gyrokinetic version, Eq.~\ref{eq:pol_gyro}, within fluid theory rather than the fluid version, Eq.~\ref{eq:pol_drift}.  That approach appears to be inconsistent because it does not carefully account for second order terms in the definition of the reference equilibrium distribution. 

So-called ``neoclassical polarization'' effects have also been considered for ``bounce-averaged'' gyrokinetic and gyrofluid theories  \cite{Gang90pfb,Beer96pop,Rosenbluth98prl,Fong99pop,Brizard00pop,Brizard07rmp,Kagan10ppcf,Wang09pop} (a comprehensive overview of the literature is given in Ref.~\onlinecite{Wang09pop}).    
A similar paradox occurs in this context due to the finite orbit width (FOW) in the sense that the total effect is twice as large as that due to the orbit average alone. 
For example, in the derivation of the Rosenbluth-Hinton residual flow \cite{Rosenbluth98prl}, it is clear that a second spatial averaging operation must be performed in real space in order to obtain the final result.    
We plan to explore  polarization due to neoclassical effects in future work.

\subsection{Summary of Results}

The main goal of this work is to show that a careful accounting of the translation between the real phase space coordinates and the action-angle phase space coordinates resolves the \emph{diamagnetic polarization paradox}.
It may seem obvious that the resolution has something to do with the transformation, $\CT:\zcoord\rightarrow\Zcoord$, between real space, $\zcoord=\{\xcoord,\vcoord\}$, and adiabatic coordinates, $\Zcoord=\{\Rcoord,\Vpar,\Jcyc,\gyrophase\}$.  
However, since going back, $\zcoord=\CT^{-1}(\Zcoord)$, and forth, $\Zcoord=\CT(\zcoord)$, between coordinate systems is completely reversible, it is not immediately obvious where the factor of two can possibly come from.  

Polarization is measured by the displacement of a particle from a given reference position, and, hence, the polarization density is measured by comparing the actual density with a specified reference density. 
The key difference is that the standard drift theory and the standard adiabatic theory use different choices for the reference distribution functions and, hence, for the reference densities. 
In both cases, the reference distributions are considered to be the ``zeroth order'' distribution function, but the assumptions lead to different prescriptions for the zeroth order distribution.
In other words, with the standard definitions that are commonly used in the literature, {\bf the drift centers are not the same as the guiding centers!}
When a mutually consistent set of assumptions is used, the differences are only apparent and lead to the same physical results.

For the DK/DF approach, the calculations are performed in real space,  $\zcoord=\{\xcoord,\vcoord\}$, and the zeroth order particle distribution function (PDF), $\pdfstart(\xcoord,\vcoord)$, which defines the zeroth order density, $\densitystart(\xcoord)$, is specified in the limit of vanishing gyroradius, $\kperp\gyro\rightarrow 0$; i.e. infinite magnetic field strength.
In contrast, for the GC/GK/GF approach, the calculations are performed in action-angle coordinates $\Zcoord=\{\Rcoord,\Vpar,\Jcyc,\gyrophase\}$
and the zeroth order PDF, $\PDF(\Rcoord,\Vpar,\Jcyc)$, which defines the guiding center density, $\Density(\Rcoord)$, is presumed to be a function of adiabatic coordinates, i.e. the action, $ \Jcyc\simeq \mass\mu/\charge =\mass\vperp^2/2\Omgc$, the parallel velocity, $\Vpar\simeq \vpar$, and the slowly varying guiding center coordinates, $\Rcoord\simeq \xcoord$, alone.
The  real space DK/DF space approach and the action-angle GC/GK/GF approach are completely consistent with one another when the same set of assumptions is used to specify the zeroth order PDF. 
Assuming that the DK/DF and GC/GK/GF perturbation series are both convergent, then they will converge on equivalent results.
The resolution to the paradox requires carefully distinguishing between the real space PDF, $\pdf(\xcoord,\vcoord)$,  the guiding center space PDF, $\PDF(\Rcoord,\Vpar,\Jcyc)$,  and the zeroth order drift PDF, $\pdfstart(\xcoord,\vpar,\mu)$, in the limit of vanishing gyroradius $\kperp\gyro\rightarrow 0$.

For both approaches, the polarization terms have the structure of a velocity moment of the difference between the response in real space and the orbit-averaged  response. 
The orbit-averaged response is determined by a gyroaverage corresponding to the true orbit average over the particle trajectory (defined Appendix \ref{sec:gyroaverage}).
In order to return to real space, one then performs an average over the perpendicular velocity angle in real space (defined in Appendix \ref{sec:vspace-avg}).
This second velocity moment operation is similar to taking a second gyroaverage. 
It is actually important to distinguish between these two types of averaging operation to obtain the correct results.

In principle, it is straightforward to work to first order in amplitude in a spatially varying $  \VB(\xcoord)$, e.g. as in standard GC/GK theory or as in Ref.~ \cite{Parra11ppcf}.
However, doing so requires confronting the \emph{Spitzer paradox:} the fact that the single particle guiding center drifts are not the same as the collective fluid drifts.
The resolution to this paradox is simply that the true particle velocity is the sum of the single particle drift and the rotation around the guiding center.
While many authors recognize the need for the so-called magnetization law to hold for the current (see e.g. \cite{Hazeltine92book, Brizard13pop, Sugama22pop, Brizard24jpp}), it does not seem to be widely appreciated that there is a direct corollary that we call the \emph{Spitzer polarization paradox}: the single particle polarization drifts are not the same as the collective fluid polarization drifts.
The resolution is similar:  the rate of change of the magnetization current induces a polarization current that is not divergence-free, and, hence, {\bf in a magnetized plasma, a  magnetization current that changes in time induces a nontrivial polarization charge density.}
In Part II \cite{Joseph2025dkgc-II}, we prove that a slight reinterpretation of the calculations of Brizard \& Tronko \cite{Brizard16arxiv} resolves the Spitzer polarization paradox within symplectic guiding center perturbation theory.

To keep the discussion as  simple and clear as possible, for much of our discussion of gyrokinetics, the magnetic field, $\VB$, will be assumed to be  constant in both magnitude and direction.  In a constant magnetic field, the magnetic drifts precisely vanish and the magnetization current  
is equal to the diamagnetic current.  In a spatially varying $\VB$ field, the magnetic drift effects are subdominant because they are proportional to the temperature anisotropy and are smaller by the ratio of the pressure gradient scale length to the magnetic field scale length.  Thus, the constant $\VB$ assumption simplifies the discussion and eliminates  potential confusion regarding the need to retain $\Vnabla \VB$ effects \cite{Parra11ppcf}. 
In Part II, we will show how issues with spatial variation of the magnetic field are resolved within guiding center theory.

Collisional kinetic equilibria can only have strong dependence on local constants of the motion.
Yet, constants of the motion that are local in space and time do not generate finite orbit width effects. 
Hence, equilibria that depend only on the Hamiltonian or toroidal momentum
do not generate net polarization: the electric and thermodynamic polarizations must precisely cancel. 
In contrast, a collisionless equilibrium can have anisotropic dependance on the magnetic moment which generates a net polarization proportional to the temperature anisotropy.
It is shown that the typical form of a collisionless equilibrium is actually a ratio of gyroaverages and the size of the gyrovarying corrections are estimated.

\subsection{Outline}
The next section discusses the framework for calculating the polarization charge density.
The diamagnetic polarization paradox and the resolution to the paradox are explained in Sec.~\ref{sec:diamag_paradox}, while the detailed derivations are presented in the remainder of the paper.   
Drift theory, presented in Sec.~\ref{sec:real_space}, considers adiabatic theory in real space and 
drift fluid (DF) theory, presented in Sec.~\ref{sec:drift-fluid}, provides the quickest route to deriving polarization and magnetization in terms of velocity-space moments of the particle distribution function (PDF).
Drift kinetic (DK) theory, presented in Sec.~\ref{sec:drift-kinetic}, is required to understand the form of the zeroth order reference distribution. 
When the results are expressed in moments of the PDF, it ultimately yields the same results as fluid theory. 

Gyrokinetic (GK) theory is presented in Sec.~\ref{sec:gk} and the derivation of the GK polarization to first order in perturbation amplitude and all orders in $\kperp\gyro$ is given in Sec.~\ref{sec:gk_linear} for constant $\VB$.
Applications to the anisotropic Maxwellian and Maxwell-Boltzman distributions correct to all orders in $\kperp \gyro$ are given in Sec.~\ref{sec:applications}. 
The effects of collisions, sources, and nonlinearities are considered in Sec.~\ref{sec:kinetic}.
Conclusions are summarized in the final section.

The coordinate transformations needed to derive the results are presented in the appendices so that  the different levels of description are easily compared to each other.
The zeroth order guiding center transformation is given in Appendix~\ref{sec:gc_coordinates},
the velocity space angle average is defined in Appendix~\ref{sec:vspace-avg}, and the gyroaverage is defined in Appendix~\ref{sec:gyroaverage}.
The gyrokinetic coordinate transformation is defined in Appendix~\ref{sec:gk-transformation}.
The gyrokinetic transformation correct to first order in amplitude is used in Sec.~\ref{sec:gk_trans} 
to derive electric polarization effects.

\section{ Polarization  \label{sec:pol}}

\subsection{ Maxwell's Equations in Dielectric Media  \label{sec:maxwell_dielectric}}
Maxwell's equations determine the electric field, $\VE$, and magnetic field, $\VB$, in terms of the total electric charge density, $\chargedensity$, and total electric current density, $\VJ$ (assuming that there are no magnetic monopoles and, hence, no corresponding magnetic charge or current density).
The dynamic Maxwell equations consist of {\it Ampere's Law} and  {\it Faraday's Law} 
\begin{align}
 \partial_\tcoord \varepsilon_0\VE+ \VJ &= \nabla\times\VB /\mu_0 & \partial_\tcoord \VB &=-\nabla \times \VE 
\end{align}
where $\varepsilon_0$ is the electric permittivity  in vacuum and $\mu_0$ is the magnetic permeability in vacuum.
The  Maxwell constraint equations are simply the statements of {\it Gauss's law} (which leads to the \emph{Poisson equation}),
\begin{align}
 \nabla\cdot  \varepsilon_0 \VE&= \chargedensity  &\nabla\cdot \VB &=0  
\end{align}
that are implied by charge conservation
\begin{align}
\partial_\tcoord \chargedensity+\nabla\cdot\VJ=0.
\end{align}

In a dielectric medium, Maxwell's equations determine the electric displacement field, $\VD$, and the magnetizing field, $\VH$
in terms of the free charge density, $\chargedensity_\free$, and the free current density, $\VJ_\free$.
The dynamic Maxwell equations now modify {\it Ampere's Law} while retaining {\it Faraday's Law} 
\begin{align}
 \partial_\tcoord \VD+ \VJ_\free &= \nabla\times\VH  & \partial_\tcoord \VB &=-\nabla \times \VE .
\end{align}
Maxwell's constraint equations modify {\it Gauss's law} to
\begin{align}
 \nabla\cdot  \VD&= \chargedensity_\free &\nabla\cdot \VB &=0  .
\end{align}
These equations must be closed with  constitutive relations 
\begin{align}
\VD&=\varepsilon_0 \VE+\VPol &  \VH&= \VB/\mu_0 + \nabla \times  \VMag.
\end{align}
that specify the electric displacement, $\VD[\VE,\VB]$, in terms of the electric polarization density, $\VPol[\VE,\VB]$, and the magnetizing field, $\VH[\VE,\VB]$, in terms of the magnetization, $\VMag[\VE,\VB]$. These are assumed to be (potentially) nonlinear integro-differential functionals of the fields.

The total charge density and current density are defined as the sum of free and bound contributions, respectively
\begin{align}
\chargedensity&=\chargedensity_\bound +\chargedensity_\free& \VJ=\VJ_\bound+\VJ_\free.
\end{align}
Consistency with Maxwell's equations require the polarization and magnetization to specify the bound charge density, $\chargedensity_\bound$, and bound current density, $\VJ_\bound$, via
\begin{align}
 \chargedensity_\bound&=-\nabla\cdot\VPol &  \VJ_\bound &=\partial_\tcoord\VPol + \nabla\times\VMag.
\end{align}
The modified Gauss' law implies charge conservation for the bound and free charge densities independently, so that
\begin{align}
	 \partial_\tcoord \chargedensity_{ \free} +\nabla \cdot \VJ_{ \free}=-\partial_\tcoord \chargedensity_{\bound} -\nabla \cdot \VJ_{\bound }=0  .
\end{align}
\emph{In general, there is not a unique splitting of the total charge and current densities into bound and free components.}
One reasonable assumption for a natural splitting is to require that $\VPol$ and $\VMag$ must remain bounded in time.

The total current is the sum, $\VJ=\VJ_\pol +\VJ_\mag$, of the polarization current and magnetization current, defined as
\begin{align}
\VJ_\pol &=\partial_\tcoord\VPol && \VJ_\mag= \nabla\times\VMag.
\end{align}
The polarization charge density that is required to maintain charge conservation for a given polarization current is precisely the the bound charge density.
\begin{align}
\chargedensity_\pol =\chargedensity_\bound.
\end{align}
This is the relation commonly used in plasma theory.

Note, however, that the polarization/magnetization splitting is not necessarily unique because one can always add and subtract a divergence free current that is also a partial derivative in time so that 
\begin{align}
\VPol'&=\VPol - \nabla \times \VL & \VMag'&=\VMag +\partial_\tcoord  \VL  .
\end{align}
The splitting can be made unique by using Noether's theorem to define the polarization and magnetization densities \cite{Brizard13pop,Brizard2020pop}.
With this definition, the term $\nabla\times \partial_\tcoord\VL$ is considered a magnetization current because it does not couple to the electric field.

\subsection{ Polarization as a Displacement of Charges \label{sec:charge_displacement} }
 \def\VpolF{{ \Vpol_{\kin}}}
 \def\VdF{{ \Vd_{\kin}}}
 
The  electric polarization density, $\VPol$, is defined as the density of  electric dipoles, $\Vdipole$,  in a given region of space; i.e. $\VPol=\density_\dipole\Vdipole$, where $\density_\dipole$ is the number density of dipoles.
In many cases of interest, including the physics of plasmas, the electric dipole moments are defined through the displacement of the true charge density, 
$\chargedensity=\charge\density$, from the bound charge density, $\chargedensity_\bound=\charge\density_\bound$, in response to external forces.
\begin{align} \label{eq:polarization_charge}
  \chargedensity_\bound =\chargedensity- \chargedensity_\free  =- \Vnabla\cdot ( \density \Vdipole )
=- \Vnabla \cdot \VPol.
\end{align} 
 The microscopic polarization and  displacement vectors are defined through the displacement of the microscopic charges via $\Vpol=\charge\Vd$. 
 The total polarization is the sum of the microscopic polarization, $\Vpol=\charge\Vd$, and the collective polarization, $\VpolF=\charge\VdF$, which depends on gradients of the PDF of the charge distribution.
 Hence, the total polarization is
 \begin{align}
 \Vdipole=\Vpol+\VpolF  =\charge \left(\Vd+\VdF \right).
 \end{align}

Like any function in three dimensions, the charge density can be expanded in multipoles.
Assuming that the charge density is bounded in magnitude and localized in space, the (un-normalized) multipoles, defined with respect to the specific point of origin $\Vx$,  are given by \cite{Jackson99book}
\begin{align}
\Squad_{ijk\dots} (\Vx)  &=\int d^3\xcoord'   \chargedensity(\Vx')  \Delta \xcoord_i \Delta \xcoord_j \Delta \xcoord_k \cdots  
\end{align}
where $\Delta\Vx=\Vx'-\Vx$ is the relative displacement from the origin
The electrostatic energy can be expressed in terms of multipoles by Taylor expanding the electric potential around the origin, $\Vx$,
\cite{Jackson99book}
\begin{align}
W_E=\int d^3\xcoord \chargedensity \potential   &=\left( \charge  + \Vdipole_1 \cdot\nabla  + \tfrac{1}{2} \Vquad_2:\nabla\nabla + \dots\right) \potential  (\Vx)
\end{align}
where $\Vpol_0$ is the electric dipole and  $ \Vquad_0$ is the electric quadrupole moment near the origin, $\Vx$.
Next, in addition to the free charge density, one can consider independent densities of multipoles, e.g. the dipole density, $ \VPol_1=\density_1 \Vdipole_1$, and quadrupole density, $\VQuad_2 = \density_2 \Vquad_2$, which yields the expansion
\begin{align}
W_E=\int d^3\xcoord \left( \chargedensity_\free +  \VPol_1 \cdot \nabla+ \tfrac{1}{2} \VQuad_2:\nabla\nabla + \dots\right) \potential  (\Vx).
\end{align}
Finally, integration by parts  yields an expression for the charge density in terms of multipole charge densities
\begin{align} \label{eq:multipole_charge_density}
\chargedensity_\bound   = 
-  \nabla\cdot \VPol &=  \nabla\cdot \left[ -\VPol_1 + \tfrac{1}{2}  \nabla\cdot \VQuad_2 - \dots\right] .
\end{align}
Thus, one finds that
\begin{align}   \label{eq:multipole_Pvec}
 \VPol &= \density\Vdipole = \VPol_1 - \tfrac{1}{2}  \nabla\cdot \VQuad_2 + \dots
 \end{align}
Both the microscopic and macroscopic polarization have the same form as Eq.~\ref{eq:multipole_Pvec}.
While it should be obvious, we point out that since the fields are defined in real space, all gradients here are in real space $\Vx$.

The polarization density expressions for magnetized plasmas in Eqs.~\ref{eq:pol_drift} and \ref{eq:pol_gyro} have simple interpretations.
The electric polarization is due to the displacement of the gyrocenter caused by a gradient in the electric field.
Similarly, the diamagnetic polarization is due to the effective displacement generated by a pressure gradient.
In both cases, the dipole moment is generated by displacements that are second order in gyroradius; i.e. due to the contribution of the quadrupole moment in the plane perpendicular to $\VB$ to the dipole moment.

\subsection{ Quasineutrality \label{sec:quasineutrality}}

The electric field can be determined from Gauss's law
\begin{align}\label{eq:poisson}
\Vnabla\cdot \epsilon_0 \VE = \chargedensity = \sum_s\charge_s \density_s
\end{align}
which involves a sum of the charge densities, $\chargedensity_s=\charge_s \density_s$,  for each species, $s$.
However, the vacuum polarization term, $\Vnabla\cdot \epsilon_0 \VE$, generates electron plasma wave dynamics that is much faster than the timescales of interest in the adiabatic theory.  Therefore, it is more expedient to use the quasineutral approximation  
\begin{align}\label{eq:quasineutrality}
0\simeq   \chargedensity= \sum_s\charge_s \density_s.
\end{align}
In an approximately quasineutral system, the total charge density is much smaller than the bound and free charge densities independently.
Hence, the assumption of \emph{quasineutrality} implies that the  bound charge density and the  free charge density are equal in magnitude and opposite in sign, so that the total charge density nearly vanishes
\begin{align}
	 0 \simeq\chargedensity&=\chargedensity_\bound +\chargedensity_\free .
\end{align}
This can be expressed in the equivalent form of quasineutral charge conservation 
\begin{align}
	0 &\simeq -\partial_\tcoord\chargedensity=-\partial_\tcoord( \chargedensity_\bound +\chargedensity_\free)  \\
	&= \nabla\cdot\VJ = \nabla\cdot(\VJ_\bound+ \VJ_\free )
\end{align}
which can also be expressed as
\begin{align}
	 \partial_\tcoord \chargedensity_\bound = -\nabla\cdot\VJ_\bound \simeq \nabla\cdot \VJ_\free = -\partial_\tcoord \chargedensity_\free  .
\end{align}

Thus, in order to determine $\VE$ one can  either use the  \emph{quasineutral Gauss's law (Poisson equation)}  
\begin{align}
	 \chargedensity_\bound:= - \nabla\cdot \VPol[\VE,\VB] \simeq -\chargedensity_\free  
\end{align}
or 
\emph{quasineutral charge conservation}  
\begin{align}
	 \partial_\tcoord \chargedensity_\bound:=-  \partial_\tcoord  \nabla\cdot\VPol[\VE,\VB]\simeq  \nabla\cdot \VJ_\free .  
\end{align}

\section{ The Diamagnetic Polarization Paradox \label{sec:diamag_paradox} }

\subsection{The  Bound/Free Splitting is Not Unique }
The charge carriers explore the accessible phase space, $\zcoord=\{\xcoord,\vcoord\}$, which is specified in terms of the position, $\Vx$, and velocity, $\Vv=d\Vx/d\tcoord$, of each particle.
The probability distribution function, $\pdf(\tcoord,\xcoord,\vcoord)$ specifies the true particle location and satisfies the conservation of probability 
\begin{align}
0=d\pdf/d\tcoord&=\partial_\tcoord\pdf +\partial_\zcoord(\pdf \dot\zcoord)\\
&=
\partial_\tcoord\pdf +\nabla_\xcoord \cdot ( \pdf \dot \Vx) + \nabla_\Vv\cdot (\pdf \dot \Vv).
\end{align}

Now, assume  that an adiabatic coordinate transformation, $\CT:\zcoord\rightarrow\Zcoord=\{\Rcoord,\Vpar,\Jcyc,\gyrophase\}$,  can be found, where the motion can be cleanly separated into fast and slow motions.
Hence, assume the form
\begin{align}
\Vx&=\VR+\Vgyro(\tcoord,\Rcoord,\Vpar,\Jcyc,\gyrophase) \\
\Vv&=\VV(\tcoord,\Rcoord,\Vpar,\Jcyc)+\Vw(\tcoord,\Rcoord,\Vpar,\Jcyc,\gyrophase)
\end{align}
where the fast small scale motion is defined by $\Vgyro$ and $\Vw:=d\Vgyro/d\tcoord$ and the slow large scale motion is defined by $\VR$ and $\VV:=d\VR/d\tcoord$.
The action will be assumed to be constant in time to the order of the calculation, $\dot\Jcyc=0$, while motion of the associated gyrophase, $\dot\gyrophase=\Omgc$, sets the fast gyrofrequency.
The PDF in the new coordinates, $\PDF(\tcoord,\Rcoord,\Vpar,\Jcyc,\gyrophase)$, satisfies the conservation law
\begin{align}
0&=d\PDF/d\tcoord=\partial_\tcoord\PDF +\partial_\Zcoord(\PDF \dot\Zcoord)\\
&=
\partial_\tcoord\PDF +\nabla_\Rcoord \cdot( \PDF  \dot \VR)+ \partial_{\Vpar}(\PDF \dot \Vpar)+\partial_\gyrophase(\PDF\dot\gyrophase) .
\end{align}
The zeroth order GC transformation is described in Appendix~\ref{sec:gc_coordinates}.
The GK transformation is described in Sec.~\ref{sec:gk_trans}.

Now, the polarization and magnetization vectors can certainly be calculated uniquely.
However, the assumptions that go into the specification of the separation into the fast small scale motion, $\Vgyro$, and the slow large scale motion, $\VR$,  play an important role in the final answer.
Considering that this separation is usually accomplished using perturbation series with differing ordering assumptions that generically only converge  asymptotically, there is not necessarily a unique result.
In fact, there are potentially multiple issues for two different splittings, $\Vx=\VR+\Vgyro$ and $\Vx=\VR'+\Vgyro'$.
Important questions to ask are:
\begin{itemize}
\item Are the perturbation series the same or different up to some order?
\item Are the coordinates, $\VR, \Vgyro$,  the same up to some order?
\item Are the velocities, $\VV,\Vw$,  the same up to some order?
\item Are the initial PDFs the same up to some order?
\item {\bf Are the PDFs being compared to the same reference distribution, $\pdf_0$, up to some order?}
\end{itemize}
We will prove that it is this last point that explains the difference between the calculation of the polarization density in DK/DF theory and GC/GK/GF theory.
The standard practice in GC/GK/GF theory is to compare the true PDF, $\pdf(\xcoord,\vcoord)$, to the PDF in action-angle coordinates but where the true particle coordinates are used in place of the guiding center coordinates, $ \PDF(\Rcoord\rightarrow \xcoord,\wcoord\rightarrow\vcoord)  $.
In DK/DF theory, one compares the true PDF, $\pdf(\xcoord,\vcoord)$, to the PDF in the limit of vanishing gyroradius, $\pdf_0(\xcoord,\vcoord)$.
Both concepts are correct and useful within the respective theories.
However, the two categories of theories must be compared to one another carefully in order to provide the correct translation between the two pictures.

\subsection{Differences in Ordering Assumptions  \label{sec:ordering}}

One of the key issues that must be understood in comparing the results is that  the formal ordering in small parameters used by gyrokinetic (GK) and gyrofluid (GF) theory and drift-reduced kinetic (DK) and fluid (DF) theory have some important differences.  
Let $\epsnonlin$ parameterize the characteristic perturbation amplitude of the nonlinearity and let $\epsperp$ and $\epspar$ parameterize the  size of perpendicular and parallel FLR effects:
\begin{align}
\epsnonlin&\sim \charge \phipert/\Temp_\eq \sim \densitypert/\density_\eq \sim \Temppert/\Temp_\eq\\
\epsperp&\sim\kperp \gyro \\
\epspar&\sim \kpar \gyro \sim \omega/\Omgc 
\end{align} 
where $\Vk=k_\|\bhat + \Vk_\perp$ is a characteristic wavevector and $\omega$ is a characteristic frequency.
Here, $\density_\eq$ and $\Temp_\eq$ are the equilibrium density and temperature, while   $\delta \density$, $\delta \Temp$,  $\delta \phi$, are the perturbed density, temperature, and electric potential.

For any quantity, $\pdf$, one can perform the formal expansion in small parameters 
\begin{align}
\pdf = \sum_{ijk} \epsnonlin^i \epsperp^j \epspar^k  \pdf_{ijk} =\sum_{jk} \epsperp^j \epspar^k \left( \pdf_{0,jk}+\pdfpert_{jk} +\dots \right) .
\end{align}
GK/GF  theory assumes the ordering
\begin{align}
   \omega/\Omgc\sim \kpar\gyro \ll    \kperp \gyro \sim 1
\end{align}
for the frequency $\omega$ and  performs calculations to all orders in $\epsperp$.   However, it relies in an essential manner on the assumption of small amplitude, $\epsnonlin$, and small $\epspar$ in order to ensure that resonances do not destroy the adiabatic invariance of the magnetic moment. 
 In contrast,  guiding center (GC) theory,  drift-kinetic (DK) and drift-fluid (DF) theory typically assume the ordering 
\begin{align}
   \omega/\Omgc\sim \kpar\gyro \lesssim    \kperp \gyro \ll 1.
\end{align}
Here, the essential perturbation parameters are $\epsperp$ and $\epspar$. While there are still restrictions on the amplitude, all orders in  $\epsnonlin$ can be solved for at each stage of the calculation. 
From the GC/DK/DF point of view, the expressions of the total polarization density in Eq.~\ref{eq:pol_drift} and Eq.~\ref{eq:pol_gyro} are accurate  to second order in $\epsperp$ and all orders in $\epsnonlin$.  
 
Another important point is that the electromagnetic fields  must be slowly varying  in the following sense. Adiabatic theory requires the dynamical frequencies to be smaller than the gyrofrequency, which rules out the presence of the compressional Alfv\'en and sound waves  propagating perpendicular to $\VBfield$. The characteristic frequency scales, $\kperp  c_A$ and $\kperp c_S$, set by the Alfv\'en speed,  $c_A$,  sound speed, $c_S\sim V_T$, and thermal speed respectively, imply that the fast and slow magnetosonic frequencies can be of the same order or larger than the gyrofrequency as $\kperp \gyro \rightarrow 1$.  As is well known, this requires the quasi-equilibrium, so-called ``Drift-Reduced MHD'' \cite{Kruger98pop}, constraints  on the inductive electric field, $\partial_t \VA_\perp \ll \Vnabla_\perp \phi$,  and on the divergence of the perpendicular flow, ${\partial_t \Vnabla \cdot \density \Vv_\perp /\density \Omgc^2 \ll 1}$. 
 Hence, the approximation 
 \begin{align} \label{eq:electrostatic} 
 \VEfield_\perp=-\Vnabla_\perp\phi
 \end{align}
  is required for consistency because $\partial_\tcoord \VA$ is one higher order in $\epspar$. Thus,  Eq.~\ref{eq:electrostatic} can be used throughout  the following.

\subsection{ Differences in the Poisson Equation  \label{sec:gyro-poisson} }

While both DK/DF and GC/GK/GF theory typically assume quasineutrality, there is  an important difference between the manner in which Eqs.~\ref{eq:pol_drift} and \ref{eq:pol_gyro} are used in the respective theories. For the DK/DF calculation, one explicitly solves for the total polarization density, while, for the GC/GK/GF calculation, one only explicitly solves for the electric polarization density.  

For DK/DF theory, the evolution of the total polarization density is explicitly calculated using the ``vorticity equation.'' 
In fluid theory, the bound polarization charge density is called the ``vorticity'' because it is approximately equal to the parallel vorticity of the main ion species, $i$, 
\begin{align}
\chargedensity_\pol \simeq \Vnabla \cdot  \mass_i \density_i \Vv_i\times  \bhat /\Omgc \simeq \bhat/\Omgc\cdot \Vnabla \times  \mass_i \density_i\Vv_i 
.
\end{align} 
The modern derivation \cite{Chang92pfb, Zeiler97pop, Xu00pop, Simakov03pop, Simakov04pop, Parra09ppcf_vorticity} of the vorticity equation proceeds through a calculation of quasineutral charge conservation
\begin{align}
0&\simeq \partial_\tcoord \chargedensity =- \Vnabla\cdot \VJ.
\end{align}
Here, the polarization charge density is defined as the bound charge density,   $\chargedensity_\pol: = \chargedensity_\bound$;
note that $\chargedensity_\pol$ is denoted as $\varpi$ in Refs.~\onlinecite{Zeiler97pop, Xu00pop, Simakov03pop, Simakov04pop}.
  The  drift current, $\VJ_\drift$, represents the transport of free charge density, which will be referred to as $\chargedensity_\drift=\chargedensity_\free$.  Thus, the quasineutrality constraint can be expressed in the equivalent form
\begin{align}
0&\simeq \partial_\tcoord \left(\chargedensity_\drift+\chargedensity_\pol \right) =- \Vnabla\cdot\left( \VJ_\drift + \VJ_\pol \right).
\end{align}
 The final step is to determine the electric potential, $\phi$,  from the free charge density, $ \chargedensity_\drift$, either through the quasineutral Poisson equation
\begin{align} \label{eq:quasineutrality_drf}
\chargedensity_\pol\simeq -\chargedensity_\drift.
\end{align}
or through the \emph{vorticity equation}, i.e., the quasineutral charge conservation law 
\begin{align} \label{eq:vorticity_eq}
 \partial_\tcoord  \chargedensity_\pol \simeq  \Vnabla\cdot  \VJ_\drift.
\end{align}
Thus, the DK/DF formalism corresponds precisely to the discussion in Sec.~\ref{sec:quasineutrality}.

In contrast, for GC/GK/GF theory, the electric potential is usually determined by directly solving the so-called \emph{gyro-Poisson equation}. This version of quasineutrality equates the polarization charge density due to the electric field alone, $\chargedensity_{\pol,\Efield}$, directly to the difference between the guiding center densities in real space, e.g.
 \begin{align} \label{eq:Epol_gyro}
   \chargedensity_{\pol,\Efield} (\xcoord)
   &=
   -\sum_s \int  \charge_s \PDF_s\delta^6(\zcoord-\CT_s^{-1}(\Zcoord))d^6\Zcoord d^3\vcoord
 \end{align}
 where the guiding center coordinates, $\VR$, are defined in Appendix~\ref{sec:gc_coordinates}.
If the drift velocity is neglected then this expression reduces to
 \begin{align} \label{eq:Epol_gyro}
   \chargedensity_{\pol,\Efield} (\xcoord)
   &=
   -\sum_s \int  \charge_s  \PDF_s\delta^3(\Vx-\VR-\Vgyro(\Zcoord))d^6\Zcoord
   .
 \end{align}
 The diamagnetic polarization terms are implicit in the expression on the right hand side and it is clear that these terms must be due to finite Larmor radius effects. 
Thus, another important difference is that, in the GC/GK/GF formulation, the total real space polarization density, $\chargedensity_\pol=\chargedensity_{\pol,\Efield}+\chargedensity_{\pol,*}$, is never explicitly calculated.  

Finally, an accurate expression for the density of each species is a nonlinear integro-differential functional of the electric and magnetic fields, $\density_s[\VE,\VB;\xcoord]$.  This makes solving either form of the Poisson equation, Eq.~\ref{eq:poisson} or the quasineutrality condition, Eq.~\ref{eq:quasineutrality}, rather difficult. Thus, it is common to derive the quasineutrality condition to first order in the amplitude of the potential, because in this case, the electric polarization can be cleanly separated from the diamagnetic polarization and results in the vorticity and gyro-Poisson equations. 

\begin{figure}
\includegraphics[width=2in]{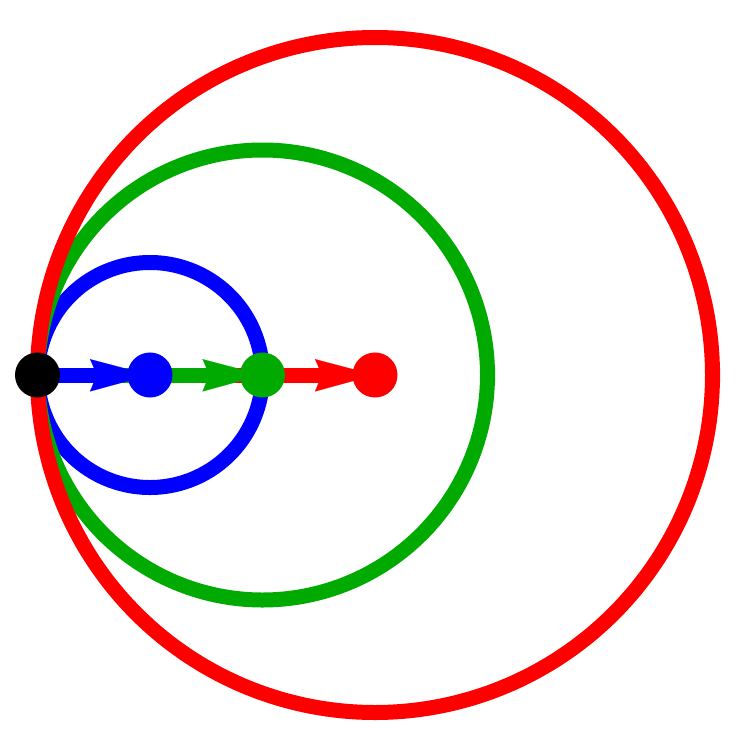}
\caption{At constant particle position (black dot), an increase in gyroradius, illustrated by a change in color from blue to green to red, causes a shift in the guiding center position (colored dots) indicated by arrows. }
\label{fig:GyroShift}
\end{figure}

\subsection{Resolution of the Diamagnetic Polarization Paradox  \label{sec:resolution} }

The paradox  is due to the difference between the density of particles in real space, $\density(\xcoord)$,  and the density of guiding centers in guiding center space, $\Density(\Rcoord)$,  often called the ``guiding center'' or ``gyrofluid'' density.
 GC/GK/GF theory gives a definite prescription for  mapping the guiding center density to the real space density, $\Density(\Rcoord)\rightarrow \density(\xcoord)$.  
However,  DK/DF theory begins with a different starting point: the fluid density, $\densitystart(\xcoord)$, in the limit of vanishing FLR effects, $\kcoord\gyro\rightarrow 0$.
In order to compare the two approaches, one must also describe the mapping $\densitystart(\xcoord) \rightarrow\Density(\Rcoord)$. 
Understanding this relationship provides the resolution to the paradox.

There is an intuitive explanation of the need to understand this mapping \cite{Bolton19pvt}.  Imagine adiabatically increasing the perpendicular velocity, $\vperp$, of a particle without changing its position. Clearly, this will generate an increase in the gyroradius, $\gyro=\vperp/\Omgc$, proportional to the increase in $\vperp$. However, in addition to increasing the size of the orbit, it also causes the guiding center position to be displaced from the particle position by the increase in gyroradius, as shown in Fig.~\ref{fig:GyroShift}.   

\begin{figure}[t]
\includegraphics[width=2.5in]{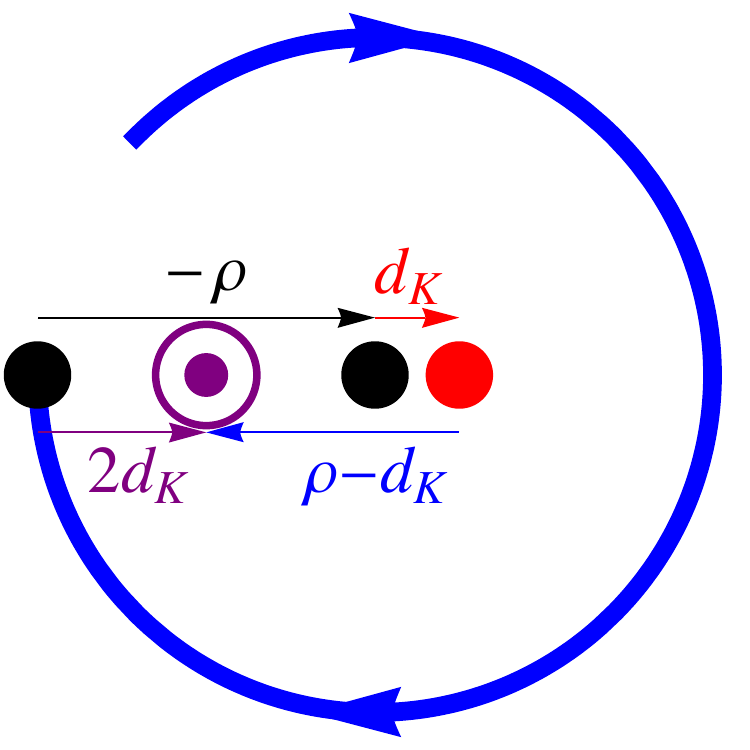}
\caption{For drift theory, the two stages of FLR effects double the \kinetic polarization, $\VpolF=\charge \VdF$, driven by gradients of the PDF. In the figure, the pressure gradient increases to the left.
 (1) The first stage displaces the zeroth order real space particle position (black circle) by the amount $-\Vgyro$ (black arrow) to the guiding center position (black).
 The pressure gradient shifts the polarization center further out by $\VdF=\gyro_T^2/2L_\press$   (red arrow) to the (red disk), 
 where $1/L_\press=\left|\nabla_\perp \ln{\press}\right|$.
(2) The second stage displaces the polarization center (red disk) back by the amount $\Vgyro$, but the shift including the pressure gradient is reduced to $\Vgyro-\VdF$ (blue arrow), which yields the final polarization center (purple disk) in real space.
This is equivalent to displacing the zeroth order particle position by $2\VdF=\gyro_T^2/L_\press$ (purple arrow, purple circle). 
Guiding center theory only has one stage, so the total pressure-driven displacement is $\VdF=\gyro_T^2/2L_\press$.}
\label{fig:GyroPol}
\end{figure}

Thus, there are two steps required to describe  diamagnetic polarization. 
First, ``turn on'' the finite Larmor radius effects via the transformation $\densitystart(\xcoord) \rightarrow\Density(\Rcoord)$. 
Second, map the GK distribution back to real space via the coordinate transformation $\Density(\Rcoord)\rightarrow \density(\xcoord)$.  
As illustrated in Fig.~\ref{fig:GyroPol}, the effect of the displacement between the two coordinate systems cancels at first order in gyroradius (referred to as $\hat \gyro$ in Fig.~\ref{fig:GyroPol}), 
but coherently sums at second order in gyroradius (referred to as $\bar \gyro$ in Fig.~\ref{fig:GyroPol}.
To first order in amplitude, the net effect of each of the two stages, $\densitystart(\xcoord) \rightarrow\Density(\Rcoord)\rightarrow \density(\xcoord)$,  can be expressed in terms of ``gyroaverages" and are equal in magnitude.

If both Eqs.~\ref{eq:pol_drift} and \ref{eq:pol_gyro} are true, then the real space density can be expressed in two different ways: 
\begin{align}
\density&\simeq\densitystart  +  \Vnabla\cdot   \left[\charge\densitystart   \Vnabla_\perp \phi +\left(\Vnabla\cdot \pressTensor _{\zeroflr}  \right)_\perp \right] /\mass\Omgc^2\\
&\simeq\Density +  \Vnabla\cdot   \left[ \charge\Density \Vnabla_\perp \phi + \tfrac{1}{2} \left(\Vnabla\cdot \PressTensor \right)_\perp\right] /\mass\Omgc^2.
\label{eq:density_realspace}
\end{align}
This can only be true if the guiding center density is related to the zeroth order density via
\begin{align} \label{eq:pol_gyro-drift}
\Density&\simeq \densitystart  + \tfrac{1}{2} \Vnabla\cdot  \left(\Vnabla\cdot \pressTensor_\zeroflr \right)_\perp /\mass\Omgc^2.
\end{align}
This has the intriguing interpretation that the guiding center density, $\Density(\Rcoord)$, appears to be the ``gyroaverage'' of the zeroth order density, $\densitystart (\xcoord)$, to second order in FLR effects.
Thus,   the full diamagnetic polarization density in the real space picture should actually be considered to be 
\begin{align} 
{\rm GC/GK/GF:} && \density_{\pol}     \simeq   \left. \Vnabla  \cdot   \left[ \charge\Density \Vnabla_\perp \phi + \left(\Vnabla\cdot \PressTensor \right)_\perp\right] /\mass\Omgc^2\right|_{\xcoord}
\label{eq:pol_gyro_corrected}
\end{align}
which agrees with the DK/DF result to second order in FLR. Note that Ref.~\onlinecite{Xu13pop} used this relation as an important step in deriving their gyrofluid equations (their Eq.~15).

It is well known that the gyroaveraging operation  introduces a smearing of the distribution over the gyroradius.  For example, the real space PDF, $\pdf(\xcoord,\vcoord)$, is determined by transformation of the guiding center space PDF, $\PDF(\Rcoord,\Vpar,\Jcyc)$,  from gyrokinetic coordinates to real space coordinates,.
Since only the velocity angle average of the real space distribution $\avgVel{\pdf}$ (defined in Eq.~\ref{eq:gyroaverage_x}) contributes to the density in real space, this introduces 1/2 of the smearing effect required to derive the full polarization density.   This seems to indicate that a ``double gyroaverage'' would give the desired result.  

An example of how the guiding center position shifts as the gyroradius is increased in shown in Fig.~\ref{fig:GyroShift}.
However, if one  transforms  the   PDF in real phase space coordinates, $\pdf(\xcoord,\vcoord)$, back to adiabatic phase space coordinates, 
then one obtains precisely the original PDF, $\PDF(\Rcoord,\Vpar,\Jcyc)$, which removes all smearing effects.  Hence, the explanation must involve something in addition to the  coordinate transformation.

It will be shown that, when the DK/DF assumptions are satisfied, the PDF in gyrokinetic coordinates can be expressed as the gyroaverage of the zeroth-order distribution in real space, $\pdfstart(\xcoord,\vcoord)$ to first order in amplitude, $\epsnonlin$, and all orders in $\kperp\gyro$.
The ``zeroth order'' distribution, $\pdfstart$, is defined to be the PDF in the limit of vanishing $\kperp\gyro\rightarrow 0$. 
Crucially, $\pdf_0 $ has the same guiding center drift velocity as $\PDF$ to lowest order, but has all other finite orbit width effects removed. 
To simplify the discussion, the drifts are assumed to vanish for the remainder of this section.

The gyroaveraging operation (defined in Eq.~\ref{eq:gyroaverage_R})
 introduces a smearing  of the  real space $\pdfstart$ distribution to the form 
\begin{align}
\PDF(\Rcoord,\Vpar,\Jcyc) = \avgCyc{\pdf_0(\zcoord(\Zcoord))} \simeq\left. \pdfstart+ \tfrac{1}{4} \nabla_\perp^2 \gyro^2 \pdfstart+\dots \right|_{\Rcoord}.
\end{align}
 If  $\densitystart(\xcoord)$ is the density of $\pdfstart(\xcoord,\vpar,\mu)$ and $\Density(\Rcoord)$ is the density of $\PDF(\Rcoord,\Vpar,\Jcyc)$, then they are related by
\begin{align}
\Density(\Rcoord)  &\simeq \left. \densitystart + \tfrac{1}{2}  \nabla\cdot\nabla_\perp \pressstart \, \,   \gyro_T^2/\Tperp+\dots  \right|_{\Rcoord}.
\end{align}
Here, the thermal gyroradius is defined by $\gyro_T^2 =  \Tperp/\mass \Omgc^2$, where $\Tperp$ is the perpendicular temperature, and $\pressstart=\densitystart \Tperp$ is the perpendicular pressure in the limit of vanishing gyroradius. The difference between  $\Density(\Rcoord)$ and $\density_0(\xcoord)$ is due to the displacement of the guiding centers from their position at zero gyroradius, shown schematically in Fig.~\ref{fig:GyroPol}.

In order to determine the density in real space, one must also perform a velocity average, which is the equivalent of the gyroaveraging operation in real space (defined in Eq.~\ref{eq:gyroaverage_x}).  
Since $\pdf(\zcoord)=\PDF(\Zcoord(\zcoord))$, this leads to the gyroaveraged real space distribution 
\begin{align}
\avgVel{\pdf}(\xcoord,\vpar,\Jcyc) =\avgVel{\PDF(\zcoord(\Zcoord))}
& \simeq \left. \PDF+ \tfrac{1}{4} \nabla_\perp^2 \gyro^2\PDF+\dots  \right|_{\xcoord} \\
&\simeq \left. \pdfstart+ \tfrac{1}{2} \nabla_\perp^2\gyro^2 \pdfstart +\dots \right|_{\xcoord}.
\end{align}
The density, $ \density(\xcoord)$,  corresponding to $\pdf$ in real space is 
\begin{align}
\density (\xcoord)\simeq \left. \densitystart+   \nabla\cdot \nabla_\perp \pressstart \, \,  \gyro_T^2/\Tperp +\dots \right|_{\xcoord}.
\end{align}
The gyroaveraging operation involved in this equation leads to the factor of 1/2  in Eq.~\ref{eq:pol_gyro}.  
The difference between  $\density(\xcoord)$ and $\Density(\Rcoord)$ is due to the displacement of the real space particle positions from the guiding centers.

\begin{figure}
\includegraphics[width=2.5in]{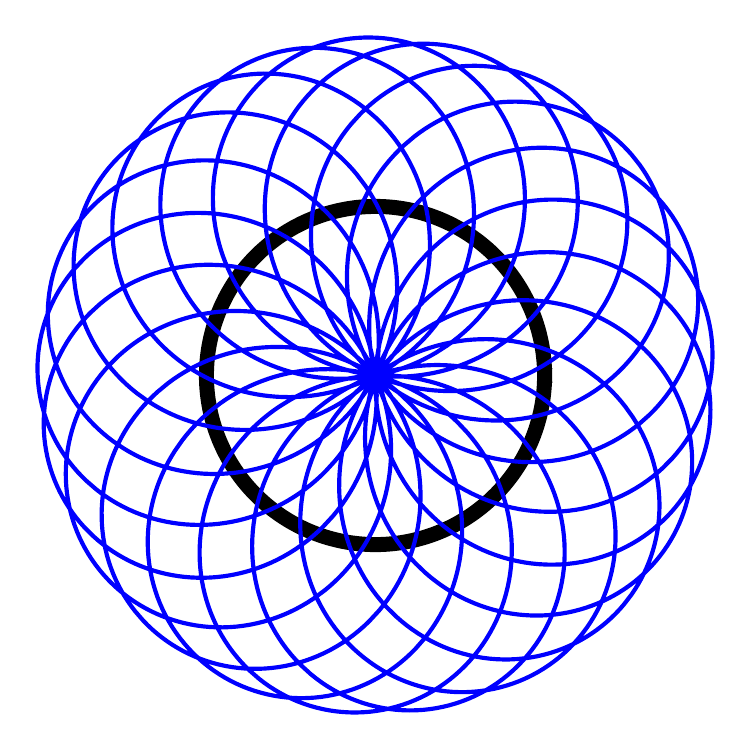}
\caption{For GC/GK/GF theory, the density in real space depends on the velocity space average of the guiding center distribution, $\avgVel{\pdf}=\avgVel{\PDF}$,  determined by the integral over the black circle.
For DK/DF theory, there are two stages of  finite orbit width effects. 
First transform $\pdf_0(\zcoord)$ to guiding center space, then gyroaverage to obtain $\PDF(\Zcoord)=\avgCyc{\pdf_0}$. 
Second,  transform  $\PDF(\Zcoord)$ back to real space, and, finally, gyroaverage to obtain the density. 
The two stage process implies that each point on the black circle is in turn defined by a gyroaverage over one of the blue circles.
While the outer radius of the blue region is twice as large,  the orbit integral over the blue region is determined by the root mean-square radius and yields twice the area of the black circle.  }
\label{fig:DriftPol}
\end{figure}

For GC/GK/GF theory, the PDF in real space is determined by the velocity space average of the PDF in guiding center coordinates, $\pdf(\zcoord)=\avgVel{\PDF}$.
As Fig.~\ref{fig:DriftPol} illustrates, this is the integral over the black circle.
However, for DK/DF theory, the PDF in guiding center space is determined by the gyroaverage over the zeroth order PDF, 
$\PDF(\Zcoord)=\avgCyc{\pdf_0}$.
Hence, each point on the black circle is determined by an average over one of the blue circles.
While the outer radius of the blue circles is twice as large, the double average weights the interior more strongly
Because the two averages are incoherent, the areas simply sum together.
Thus, each gyroaveraging operation in the two transformations contributes a factor of 1/2 and the two together lead to a factor of unity.

Expressions for the GK polarization that use the drift kinetic assumption of $\PDF=\avgCyc{\pdf_0}$  are derived in Secs. \ref{sec:gk_linear} and \ref{sec:applications}.
The results are accurate to first order in   amplitude $\epsnonlin$ and $\epspar\sim \omega/\Omgc\sim \kpar\gyro$ and  to all orders in $ \kperp\gyro$
From the DK/DF point of view, the polarization should be a difference between densities in real space, not between densities in different spaces. 
 Thus, if the diamagnetic polarization density is reported using Eq.~\ref{eq:pol_gyro} rather than Eq.~\ref{eq:pol_gyro_corrected}, then this does not yield the full real space diamagnetic polarization density, because it is missing an equally important contribution from $\Density-\densitystart$. 
   Nonetheless, reporting the DK/DF polarization density is typically a side calculation that does not impact the final GC/GK/GF results.
Since the ``zeroth order'' density, $\densitystart$, is not used or defined  in  GC/GK/GF calculations, the  diamagnetic polarization density according to the DK/DF approach is not  calculated. 

\emph{
It is important to emphasize that, when the assumptions for the zeroth order reference distribution are consistent, then the gyrokinetic and gyrofluid calculations agree in their predictions of the final result for the real space density.   
}
This is because the real space particle density is correctly predicted by Eq.~\ref{eq:density_realspace}.
Note the surprising conclusion that, if one assumes that the zeroth order charge density vanishes  in order to satisfy quasineutrality, then the sum of the guiding center charge densities only vanishes through first order and does not vanish to second order. 
In this case, consistency requires
\begin{align}
 \sum_s \charge_s \density_{0s}=0 && \longleftrightarrow & & \sum_s \charge_s \Density_{0s}\simeq   \chargedensity_{\pol,\diamag}/2 
 \end{align}
 so that {\bf the sum of the guiding center charge densities must be half the
 magnitude of the diamagnetic polarization charge density.}

 For example, as proven in Sec.~\ref{sec:applications},  the net polarization charge density must vanish  for the isotropic Maxwell-Boltzmann distribution, which leads to the second order accurate expression
\begin{align}
0 = \sum_s \chargedensity_{s} &=  \sum_s \chargedensity_{\pol ,\kin, s} + \chargedensity_{\pol, \Efield, s}   \\
&\simeq \sum_s\nabla\cdot   \left(\nabla_\perp \press_s- \charge_s\density_s\VE_\perp \right)/  \Omgc_s\Bfield.
 \end{align}
When expressing this relation terms of guiding center densities, one cannot eliminate the zeroth order terms.
The full relation is 
 \begin{align}
0= \sum_s \chargedensity_{s}& =  \sum_s  \chargedensity_{\pol ,\kin,s} + \chargedensity_{\pol, \Efield, s}   \\
&\simeq \sum_s\charge_s\Density_{0s}+\nabla\cdot   \left(\tfrac{1}{2} \nabla_\perp \Press_{0s}- \charge_s \Density_{0s} \VE_\perp \right)/  \Omgc_s\Bfield\\
&\simeq \sum_s \nabla\cdot  \left(  \nabla_\perp \Press_{0s}-\charge_s\Density_{0s}  \VE_\perp \right)/  \Omgc_s\Bfield.
 \end{align}
 Thus, the two forms are only consistent with each other because \emph{the zeroth order guiding center charge density does not vanish to second order}.
 
 The reader should be warned that, if one assumes that the sum of the guiding center charge densities vanishes $\sum_s \charge_s\Density_{0s}=?=0$, then the real space charge density does not vanish, instead it would be -1/2 of  the magnitude of the diamagnetic polarization charge density,  $\sum_s \chargedensity_{0s}=?=-\chargedensity_{\pol,\diamag}/2$.
For example,  Refs. \cite{LeeWW17pop, LeeWW18pop, LeeWW19pop}, implicitly used this assumption, which leads to a violation of the Boltzmann relation in many of their expressions; i.e. it is off by a factor of 1/2 in Eq.~9 of \cite{LeeWW17pop}, Eq.~2 of \cite{LeeWW18pop} and Eqs. 9-10 of \cite{LeeWW19pop}.
 Because this assumption is implicit in the definition of $\Density_{0s}$, it is not possible to detect the error by simply analyzing code results.
Instead, one must use fundamental understanding of the theory to perform consistent calculations.

\subsection{  Why is gyroaveraging required?
\label{sec:why_gyroaverage}}
\subsubsection{ The gyroaverage is generated by the gyrophase constraint  
\label{sec:gyrophase-implies-gyroaverage}
}
When the magnetic moment is a good adiabatic invariant, GC/GK theory provides  great  insight into the range of allowed solutions.
GC/GK theory represents an approach to determining a successive series of phase space coordinate transformations that make the motion trivial, i.e.  that reduce the motion to action-angle coordinates where the adiabatic invariant is conserved to high accuracy.
The lowest order approximation to this coordinate system will be denoted $\Zcoord =\{\Rcoord,\Vpar, \Jcyc, \gyrophase\}$ while the 
 the most accurate GC/GK coordinate system will be denoted $\bar\Zcoord =\{\bar\Rcoord,\bar\Vpar,\bar\Jcyc,\bar\gyrophase\}$.
In the generic case, the GC/GK perturbation series are only asymptotic series \cite{Lichtenberg92book}, so the greatest accuracy occurs at some finite order, and we assume that the $\bar\Zcoord $ coordinate system above yields the highest possible accuracy.

In order to eliminate fast time dependence from the kinetic equation, one must eliminate all dependence on the gyrophase.
Formally, this occurs because the gyrofrequency is ordered as much larger than the other terms in the kinetic equation.
The solution to the kinetic equation then requires the constraint
\begin{align}\label{eq:gyrovarying_constraint}
\Omgc  \partial\bar\PDF/\partial\bar \gyrophase=0.
\end{align}
The homogeneous solution can be any function of the slow coordinates, $\bar \PDF(\bar\Rcoord, \bar \Vpar, \bar \Jcyc)$, which clearly satisfies $\partial\bar\PDF/\partial\bar \gyrophase=0$.

Now, in the zeroth order GC/GK coordinate system, the zeroth order requirement is $\partial \PDF_0/\partial  \gyrophase=0$, so while the homogeneous solution $\PDF_0(\Rcoord,  \Vpar,   \Jcyc )$ is a good approximation, it is not completely independent of the true gyrophase, $\bar\gyrophase$.
The corrections are small because they only appear through higher order terms that are only generated if there are  gradients, etc.
To find the true solution, one must add a correction term, so that the zeroth order PDF actually becomes  $\bar\PDF = \PDF_0 + \Delta \PDF_0$.
If one were to solve this in the zeroth order coordinate system, one would generate a perturbation series for $\Delta \PDF_0$.
Another approach would be to simultaneously solve the perturbation series for the new coordinates and for $\Delta \PDF_0$ at the same time.
However, because the collisionless kinetic equation is a linear first order differential operator, the most straightforward route is to first solve for the new GC/GK coordinates and then transform to the new GC/GK coordinate system.
In this case, the solution reduces to the original requirement
\begin{align}
0=\partial\bar\PDF/\partial\bar \gyrophase = \partial \PDF_0/\partial\bar \gyrophase + \partial \Delta  \PDF_0/\partial\bar \gyrophase.
\end{align}
Hence, the solution is simply
\begin{align}
\bar \PDF =\avgCycBar{\PDF_0} && \Delta \PDF_0 = -\varyCycBar{\PDF_0}.
\end{align}	
The correction, $ \Delta \PDF_0$, represents higher-order term driven by the lowest order solution, $\PDF_0$. 
The corrected solution, $\bar \PDF$, is the average of the initial guess in the final coordinate system.
Importantly, it is not the solution obtained making the straight substitution $\Zcoord\rightarrow\bar\Zcoord$,  i.e. $\bar\PDF_0\neq \PDF_0(\bar\Rcoord,\bar \Vpar, \bar \Jcyc)$.
Rather it is the solution obtained by performing the coordinate transformation and then taking the true gyroaverage with respect to $\bar\gyrophase$.

One might object that there will be many $\gyrophase$-dependent terms that make it difficult to explicitly verify that the relation above is satisfied to all orders.
Yet, one is simply representing the exact same kinetic equation in two different coordinate systems.
The projection of the kinetic equation onto Fourier harmonics of $\bar\gyrophase $ or $\gyrophase$  is simply a choice of basis.
Since the Fourier harmonics are complete for the periodic functions of interest here, the two equations have the exact same information content.

The fact that the true gyrophase constraint is correctly applied results from a homotopy argument.
The gyrophase angles can be smoothly deformed into one another as a function of angle, $\bar\gyrophase=\gyrophase+\CO(\epsilon)$.
Thus, any gyrophase-dependence must be generated by a corresponding explicit gyrophase-dependent term in both coordinate systems.
For example, homogeneous gyrophase-dependent terms of the form $\bar\gyrophase-\bar\Omgc\tcoord$ can only arise if a corresponding homogeneous term of the form $\gyrophase-\Omgc\tcoord$ is assumed for the low order solution.
Yet, by assumption, homogeneous terms of this form    are not allowed in either coordinate system.
Similarly, any gyrophase-dependent sources must appear explicitly in both coordinate systems. 
However, by assumption, there are no gyrophase dependent sources.
The effect of sources and nonlinearity, etc., will be considered in Sec.~\ref{sec:kinetic}.

While both $\PDF_0$ and $\bar \PDF$ are reasonable reference distributions, according to GC/GK theory, the most accurate reference distribution is $\bar \PDF$ itself.
Thus, the initial approximation $\PDF_0$ should be corrected to $\bar\PDF$ by carrying the gyrophase constraint
to higher order in FLR effects.
Yet, if one begins with the initial guess $\PDF_0$, then the actual reference distribution that will be generated is precisely $\avgCycBar{\PDF_0} $.

A similar process happens in drift kinetic (DK) theory 
\cite{Hazeltine1973pp, Hazeltine92book, Helander02book} which expresses the kinetic equation in real phase space coordinates $\zcoord=\{\xcoord,\vcoord\}$.
In this case, the velocity angle is a coordinate used to describe the phase angle of the perpendicular velocity.
One is allowed to make the initial guess  $\pdf_0(\xcoord,\vpar,\mu)$ because of the assumption that the gyroradius/gyrofrequency is much smaller/larger than the spatial/temporal scales of interest.
However, there must also be a correction, $\bar \PDF = \pdf_0+\Delta \pdf_0$, that eliminates the  dependence on gyrophase due to the fact that $\pdf_0$ depends on $\Vx$ rather than $\VRbar$.
Again, transforming to the true GK coordinate system generates the gyrophase constraint
\begin{align}\label{eq:gyrophase-constraint}
0=\partial\bar\PDF/\partial\bar \gyrophase =\left. \partial \pdf_0/\partial\bar \gyrophase \right|_\Rbar
+ \left. \partial \Delta  \pdf_0/\partial\bar \gyrophase \right|_\Rbar
\end{align}
up to a predetermined error level, i.e. $\epsilon^\porder$, which has the solution
\begin{align}
\bar \PDF =\avgCycBar{\pdf_0} && \Delta \pdf_0 = -\varyCycBar{\pdf_0}
\end{align}	
up to the same error level, i.e. $\epsilon^\porder$.
In the limit that all orders are retained in the calculation, these expressions are exact.
If one only requires solving the constraint to the accuracy of the zeroth order GC/GK coordinate system, then $\PDF=\pdf_0+\Delta \pdf_0$ where
\begin{align}
0\simeq \left. \partial \PDF/\partial \gyrophase \right|_\Rcoord = \left. \partial \pdf_0/\partial \gyrophase \right|_\Rcoord+\left.  \partial \Delta  \pdf_0/\partial \gyrophase \right|_\Rcoord.
\end{align}
Thus, one obtains the lowest order approximation
\begin{align}
  \PDF= \avgCyc{\pdf_0} && \Delta \pdf_0 =  -\varyCyc{\pdf_0}.
\end{align}

Sometimes it is not appreciated that the true DK solution must have vanishing gyrovarying part in terms of the true gyrophase (Eq~\ref{eq:gyrophase-constraint}).
However, if the velocity-angle correction terms are not added to the final result, then the DK solution will have fast temporal variation, due to the fact that the position $\Vx$ depends on gyrophase through $\Vgyro(\gyrophase-\Omgc\tcoord)$, which is not the physically correct answer.
Even if this was the physically correct answer, for many applications, the desired approximation for the long time scale behavior would be the time average   $\PDF=\avgCyc{ \pdf_0}$.

It is important to point out that this lowest order DK solution  is not the most general GC/GK solution, which is instead $\PDF_0(\Rcoord,\Vpar,\Jcyc)$, or, rather, if all higher order terms are kept,   $\avgCycBar{\PDF_0}$.
Yet, as we discuss next, it is expressive enough to be considered ``complete'' for the functions of interest which are not allowed to have strong variations on  gyroradius  length scales.

The gyroaveraged form is the correct form for the polarization if it is defined as a displacement between two densities in real space. 
This is the only possible outcome when using the DK/DF ansatz that the lowest order PDF depends on real space, $\pdf_0(\xcoord,\vpar,\mu)$, rather than guiding center space.
As we prove in Sec.~\ref{sec:real_space}, this is precisely the ansatz used by the drift kinetic theory and the drift fluid moment equations.

\subsubsection{ Completeness of the gyroaveraged form }	

The fact that the corrected zeroth order distribution must have the form of a gyroaverage is an overly restrictive requirement.
For example, the gyroaverage of a function of space alone, such as the density $\density(x)$, is simply a convolution over a smoothing operation.  In Fourier space, this can be written as the product of a Bessel function and the Fourier harmonics $\avgCyc{\density}_k=J_0(\kperp\gyro) \density_k$.
Since, $J_0(\kperp\gyro)$,  the Bessel function of order 0, vanishes at certain points, $\kperp\gyro = 2.4048, 5.5201, 8.6537, \dots$, the Fourier harmonics of the gyroaverage must also vanish at these points.
Thus, the gyroaveraging operator is not invertible and, while any DK distribution, $\pdf_0$, generates a valid GC/GK distribution through the gyroaverage, $\PDF=\avgCyc{\pdf_0}$, the inverse is not true.

Yet the completeness of the DK reference distribution, $\pdf_0$, is more than adequate to describe any $\PDF$ that satisfies the foundational assumption, $\kperp\gyro\ll 1$, i.e. that there is scale separation between the gradient length scales and the gyroradius.
The same is true if one instead uses the lowest order GC approximation, $\PDF_0$.
Since the DK and GC theories both assume that the length scales of interest are much longer than the gyroradius, the behavior for spatial scales well below the gyroradius scale is irrelevant.
As higher order correction terms are added, the corrected gyroaveraged form is generated within a given order of accuracy.

For example, the accuracy of the DK/DF approximation is quite good if one assumes that $\PDF(\Rcoord,\Vpar,\Jcyc)$ is an analytic function. 
In this case, it must be bounded by exponential decay   in Fourier space, so that
\begin{align}
\abs{\PDF_k} \leq c e^{-\abs{k \ell}}
\end{align}
for some scale length $\ell$. 
The DK/GC assumption requires $\ell\gg \gyro$, so that both $\PDF_k$ and $\avgCyc{\pdf_0}_k$ are exponentially small for $\kperp\gyro>0$, hence, different values for such high harmonics generate negligible difference in real space.
For example, if $\ell/\gyro\sim 5, 10, 15, \dots$, then the difference is less than $10^{-2},10^{-4},10^{-6},\dots$, respectively.
If the exponential decay is quadratic in Fourier space
\begin{align}
\abs{\PDF_k} \leq c e^{-(k \ell)^2}
\end{align}
then the difference is even smaller. 
Now, if $\ell/\gyro\sim 2,3,4 \dots$, then the difference is of order $10^{-2},10^{-4},10^{-7},\dots$, respectively.

The analyticity assumption certainly applies to a smooth macroscopic equilibrium plasma.
However, one may need to revisit the accuracy of this assumption for a  plasma that exhibits turbulence, because this  generates power-law behavior in Fourier space, which impacts the smoothness of the solution and, hence, the relative size of the higher harmonics.
Since turbulence in magnetized plasmas is often driven by microinstabiities with  $\kperp\gyro\sim  0.1-0.5$, then one also needs to invoke a \emph{small amplitude assumption}, which is required for gyrokinetic theory to be well-defined (see \cite{Joseph21pop}).

While the GC formulation is more accurate than the DK formulation, in scenarios with strong short wavelength activity, both theories are at the border of validity, and the GK assumptions are required instead. 
 Of course, boundary layers such as the plasma sheath and small-scale sources that occur near material walls can violate the assumptions of smoothness and scale separation and, hence, require a more complete kinetic treatment.

\subsubsection{Including dissipation, sources, and nonlinearity}

In cases where the zeroth order DK solution is not a good enough guess, one cannot rely on the homogeneous DK solution alone to provide enough accuracy.
As long as the DK/GC assumptions are strong enough to guarantee convergence, then the remedy is to carry the DK expansion to higher order in FLR and solve the kinetic equation to the same order of accuracy. 
This means that one must integrate the DK equation from an earlier time where there are no particles, $\pdf(\tcoord=0, \zcoord)=0$, while including sources, so that the PDF at a later time, $\pdf(\tcoord, \zcoord)$, is completely determined.
Assuming that the DK perturbation theory converges, this procedure will generate the correct approximation for  $\bar \PDF(\bar \Zcoord)$ to the desired order of accuracy.

If we extend the physics of the kinetic equation to include dissipation, sources, nonlinear terms,  etc., then, as discussed more fully in Sec.~\ref{sec:kinetic}, a similar result will follow.
One can perform the calculation in any coordinate system that one wishes.
Assuming that the perturbation series converges, 
the same result will be obtained, up to a given specified order of accuracy, no matter which coordinate system has been used for the calculation.
At any finite order of accuracy, there may be higher order differences between the solutions.

An important point is that the different perturbation series may very well have different regions of convergence.
This will complicate the comparison if the regions of convergence do not overlap for the desired application.
The two different approaches are only certain to agree in regions where they both converge and where the assumptions lead to consistent conclusions.

\section{Drift Theory:  Adiabatic Theory in Real Space
\label{sec:real_space}}

\subsection{Drift Ordering \label{sec:drift-ordering}}
The DK/DF results can be derived from an adiabatic expansion that assumes that the gyrofrequency is much larger than all other frequencies in the system \cite{Chang92pfb,Zeiler97pop,Xu00pop,Simakov04pop}. 
In order to simplify the notation, in this section  it is assumed that $\epsperp\sim  \epspar$, i.e. that $\kperp \gyro \sim \kpar \gyro \sim \omg/\Omgc$. The species index will be suppressed when the reference to a particular species is clear.
Thus, for any quantity $\pdf$, the FLR expansion is given by
\begin{align}
\pdf = \sum_{j} \epsilon^j  \pdf_{j} .
\end{align}
The zeroth order moments for density $\densitystart$, velocity   $\Vu_\zeroflr $, and pressure, $\press_\zeroflr $, represent the guiding center moments  in the limit of ``zero orbit width.'' Higher order terms, such as $\Vu_\firstflr $, $\density_\secondflr$,  and   $\press_\secondflr $ represent  ``polarization'' and ``magnetization'' terms due to  finite Larmor radius  effects.

Within adiabatic theory, the ``free'' and ``bound'' labels can be confusing because there is not necessarily a clean separation of terms.
Thus, in this section, the individual terms will be labeled as ``drift'' terms, which are determined by a combination of ``free'' parallel motion and ``bound'' perpendicular motion, and as ``polarization'' terms that can be written as a partial derivative in time.
In principle, any polarization vector that can be written as a curl should really be considered a magnetization current (see Sec.~\ref{sec:pol}), but since this does not generate a polarization density, it is not necessary to make this distinction here.

In the rest of this section, results will be derived to second order in $\epsilon$, which is sufficient for the purposes at hand.
Thus, charge conservation, $\partial_t\chargedensity+\nabla\cdot \VJ =0$, is satisfied through second order in FLR with the definitions
\begin{align}
\chargedensity&=\chargedensity_\zeroflr +\chargedensity_{\firstflr, \drift}+\chargedensity_{\secondflr, \drift}+\chargedensity_{\secondflr, \pol}\\
\VJ&=\VJ_{\zeroflr}+\VJ_{\firstflr, \drift}+ \VJ_{\secondflr, \drift} + \VJ_{\secondflr, \pol} .
\end{align}
  The exact particle density has polarization terms of all higher orders in the FLR expansion that can be determined in a similar fashion. Similar corrections occur for all the other moments as well.

\subsection{Drift Fluid Theory\label{sec:drift-fluid}}

Thermodynamic forces drive fluxes that can be viewed equally well in the real space DK/DF picture or in the guiding center GC/GK/GF picture.  
However, the real space moment-based approach is the simplest route to deriving the relation between the thermodynamic forces and the collective fluxes because they are defined by averaging over velocity space.

The first velocity moment of the kinetic equation yields conservation of momentum. For each charged species in a plasma, the fluid momentum equation is
\begin{align}
\partial_\tcoord\mass\density\Vu+\nabla\cdot\mass\density\Vu\Vu+\nabla\cdot\Vp=\charge\density(\VE+\Vu\times\VB)+\VC+\VS
\end{align}
where $\density$ is the number density, $\Vu$ is the fluid velocity, $\Vp$ is the pressure tensor, $\VC$ is the collisional friction force with other species, and $\VS$ represents momentum sources. 
In this equation and below, all fields are functions of real space coordinates $\Vx$ and time $\tcoord$.

In a magnetized plasma, the perpendicular drift flows at order $n$ can readily be determined by Lorentz force balance:
\begin{align}
\VF_n & = \VJ_{n  } \times \VB & \VJ_{n,\perp} &= \frac{\bhat}{\Bfield}\times \VF_n.
\end{align} 
At the next order, the inertial acceleration then generates polarization flows  
\begin{align}
\partial_\tcoord \VJ_{n } & = \VJ_{n+1} \times \Omgc \bhat  
\end{align}
and yields the current of order $n+1$
\begin{align}
 \VJ_{n+1 } &=\frac{\bhat}{\Omgc} \times \partial_\tcoord \VJ_{n,\drift} 
 = \frac{\bhat}{\Omgc}\times \partial_\tcoord \left( \frac{\bhat}{\Bfield} \times \VF_n\right)
\\
&=-\partial_\tcoord \frac{ \VF_{n,\perp}}{ \Omgc \Bfield} + \left( \frac{\bhat}{\Bfield} \times \VF_n\right)\times\partial_\tcoord \frac{\bhat}{\Omgc}.
\end{align}
Since the $\partial_\tcoord \VBfield $ terms can be written in terms of  $\nabla\times\VE$, the $n+1$ order polarization current is identified as the first term
\begin{align}
\VJ_{n+1,\pol } =\partial_\tcoord \VPol_{n+1 }=-\partial_\tcoord  (\VF_{n,\perp}/ \Omgc \Bfield)
\end{align}
which yields the polarization density
\begin{align}
 \VPol_{n+1}= -\VF_{n,\perp}/ \Omgc \Bfield
\end{align} 
and the polarization charge density 
\begin{align}
\chargedensity_{n+1, \pol} =-\Vnabla \cdot   \VPol_{n+1} = \Vnabla \cdot  \VF_{n,\perp}/ \Omgc \Bfield .
\end{align} 
Clearly, the gradient of a zeroth order potential generates a first order force, a first order drift current, and a second order polarization density.

For adiabatic fluid theory,  the perpendicular velocity is determined from the perpendicular momentum equation by neglecting the small inertial acceleration, $d/d\tcoord$, terms. For each species, 
this implies that the zeroth order perpendicular velocity vanishes and that there is a first order perpendicular drift current
of the form
\begin{align}
\VJ_{ 1,\drift, \perp} =\frac{ \bhat}{\Bfield}\times \left( \Vnabla \cdot\pressTensor_0-\charge \density_0 \VE \right)
\end{align}
where, here, $\pressTensor_0$ is the pressure tensor. 
Adding the first order  parallel current, $\Jfield_{\firstflr,\|}\bhat$, completes the first order drift, $\VJ_{\firstflr, \drift}=\Jfield_{\firstflr, \|}\bhat+ \VJ_{1,\drift,\perp} $.
Note that the other forces such as inertial and viscous forces, e.g. the ram pressure associated with the drift velocity and the gyroviscosity, are not usually added to  these expressions because they are considered to be higher order in $\epsperp$ and $\epspar$.

Working to one higher order in the momentum conservation equation yields the second order   current, $\VJ_{  \secondflr}$, which determines the evolution of the second order density perturbation, $\density_{ \secondflr}$.
This leads to the second order polarization current
\begin{align}
\VJ_{2,\pol,\perp}= \partial_\tcoord \left( \charge \density_0\VE -\Vnabla \cdot\pressTensor_0   \right)_\perp/\Omgc\Bfield
\end{align}
and the second order polarization charge density
\begin{align}
\chargedensity_{ \pol, 2} =\Vnabla\cdot  \left( \Vnabla\cdot\pressTensor_0 -\charge \density_0\VE\right)_\perp /\Omgc\Bfield.
\end{align}
Thus, one clearly identifies two contributions: an electric polarization charge density driven by the electric field
\begin{align}
\chargedensity_{ \pol,\Efield} =-\Vnabla\cdot \left(\charge \density_0 \VE_\perp/\Omgc\Bfield\right)
\end{align}
and a diamagnetic polarization charge density driven by the pressure gradient
\begin{align}
\chargedensity_{ \pol, \diamag} = \Vnabla\cdot \left[(\Vnabla\cdot \pressTensor_0)_\perp/\Omgc\Bfield  \right].
\end{align}
For a scalar pressure tensor $\pressTensor_0=\press_0 \unitTensor$, this reduces to
\begin{align}
\chargedensity_{\pol, \diamag} = \Vnabla\cdot  \left(\Vnabla_\perp \press_0/\Omgc\Bfield  \right).
\end{align}
If the pressure tensor is assumed to have the gyrotropic (CGL) form 
\begin{align}
\pressTensor= \press_\|\bhat\bhat + \press_\perp (\unitTensor-\bhat\bhat)=\pi_\|\bhat\bhat+\press_\perp\unitTensor
\end{align}
where $\pi_\|:=\press_\|-\press_\perp$, then this generates the force
\begin{align}
\nabla\cdot\pressTensor&=\nabla\cdot\left[ \press_\|\bhat\bhat + \press_\perp (\unitTensor-\bhat\bhat)\right] \\
&=\nabla\press_\perp+\pi_\|\Vkappa+\VB \nabla_\|(\pi_\|/\Bfield),
\end{align}
 where $\Vkappa=\bhat\cdot\nabla\bhat$ is the magnetic field line curvature vector.
Thus, this yields the result
\begin{align} \label{eq:polarization-chargedensity_CGL-press}
\chargedensity_{\pol, \diamag} = \Vnabla\cdot  \left[\Vnabla_\perp \press_{0,\perp}+\pi_{0,\|} \Vkappa \right]/\Omgc\Bfield .
\end{align}

The part of the polarization that is driven by pressure gradient, $\nabla \press$, will be called the \emph{diamagnetic polarization}, while the part that is driven by the electric field will be called the \emph{electric polarization}.
The combination $\nabla \press-\charge \density \VE $ is the first fhermodynamic force and must vanish in any thermodynamic equilibrium without flows.
In principle, other thermodynamics forces, such as $\nabla \Temp$, can make additional contributions.
For example, this will be the case if terms higher than second order in $\kperp\gyro$ are included or if friction forces are included in the  force balance equation above.
Hence, the full kinetic response must be considered to be the sum of the generalized \emph{\kinetic polarization}, $\chargedensity_{\pol,\kin}$, and the electric polarization, $\chargedensity_{\pol,\Efield}$.

If the other forces are of zeroth order, then they should be included in the drift current
\begin{align} \label{eq:df-Jdrift}
\VJ_{ 1,\drift, \perp} =\frac{ \bhat}{\Bfield}\times \left[(\charge \density_0 \VE+\VC_0+\VS_0)-\Vnabla \cdot(\mass\density_0 \Vu_0 \Vu_0+\pressTensor_0)  \right]
\end{align}
the  polarization current
\begin{align} \label{eq:df-Jpol}
\VJ_{ 2,\pol, \perp} =-\partial_\tcoord \left[ \Vnabla \cdot(\mass\density_0 \Vu_0 \Vu_0+\pressTensor_0) -(\charge \density_0 \VE+\VC_0+\VS_0) \right]/\Omgc\Bfield
\end{align}
and the polarization charge density
\begin{align} \label{eq:df-chargedensity-pol}
\chargedensity_{ \pol} =\nabla\cdot \left[ \Vnabla \cdot(\mass\density_0 \Vu_0 \Vu_0+\pressTensor_0) -(\charge \density_0 \VE+\VC_0+\VS_0) \right]/\Omgc\Bfield.
\end{align}
However, for simplicity, we do not discuss these other terms in detail.

 
While the polarization charge density  is second order in $\epsilon$, it can be smaller than $(\kperp\gyro)^2$. 
This is because the first order drifts are \emph{intrinsically ambipolar} in the sense that the divergence of the diamagnetic current  vanishes for constant $\VBfield$. 
Direct calculation shows that $\density_{2,\pol}$ is only non-vanishing because of the weak spatial  variation of $\VBfield$
\begin{multline}
\nabla \cdot \bhat \times  \nabla \press_\tot /\Bfield =-\bhat \cdot \nabla \press_\tot \times \nabla\Bfield^2/\Bfield^2 \\
 +\nabla \press_\tot\cdot \mu_0 \VJ /\Bfield
 .
\end{multline} 
Hence, the total polarization charge density 
is of order $  \CO(\rho^2/L_BL_p)\sim \CO(\epsilon_\perp^2 L_p/L_B)$ where $L_p$ and $L_B$ are the characteristic scale lengths of $\press$ and $\Bfield$ respectively.
In contrast, the second order drifts are not typically intrinsically ambipolar.  

The final step is to solve the quasineutrality equation $\chargedensity_{\secondflr, \pol}  = -\chargedensity_{\secondflr, \drift}$ for the electric potential $\phi$, which becomes
\begin{multline} 
\Vnabla\cdot \sum_s  \left(\charge_s\density_{\zeroflr,s}\Vnabla_\perp \phi +  \Vnabla_\perp \press_{\zeroflr\perp,s}\right)\mass_s/ \charge_s\Bfield^2 
\\
= -\sum_s\charge_s \density_{\secondflr, \drift,s} .
\end{multline}
Note that the effective electric permittivity is 
\begin{align}
\varepsilon =   \sum_s \density_{0,s} \mass_s  /\Bfield^2=   (c/c_A)^2 \varepsilon_0
\end{align} 
where $c$ is the speed of light and $c_A$ is the Alfv\'en speed. For a single ion species, the electric permittivity can also be written as $\epsilon/\epsilon_0=(\Omg_p/\Omgc)^2=(\gyro_T/\lambda_d)^2$, where $\Omg_p^2= \charge^2\density/\epsilon_0\mass $ defines the ion plasma frequency and $\lambda_d=V_T/\Omega_p$ is the ion Debye length.

\subsection{Drift Kinetic Theory \label{sec:drift-kinetic}}

The kinetic equation for the real space PDF, $\pdf(\tcoord, \xcoord,\vcoord)$ can be written as 
\begin{multline}
\CK \pdf:=\left[\partial_t + \Vnabla_x \cdot \Vv + \Vnabla_{\mass \vel} \cdot \charge (\VE +\Vv\times\VB)  \right]\pdf=C[\pdf]+\Source[\pdf]
\end{multline}
where $\CollOp$ is the collision operator and $\Source$ represents sources of particles.
To simplify the discussion, we will consider the case of  constant $\VB$ and neglect collisions and sources.
In this case, the kinetic equation reduces to 
\begin{multline}
\CK \pdf:=\left[\partial_t + \Vnabla_x \cdot \Vv + \Vnabla_{\mass \vel} \cdot \charge \VE + \Omgc\partial_\gyrophase \right]\pdf
=0
\end{multline}
where the gyrophase $\gyrophase$ is defined in Appendix~\ref{sec:gc_coordinates}.
However, we note that the effects of the collision operator can be handled naturally in the drift kinetic (DK) formalism and lead to classical and neoclassical transport.

The drift kinetic equation is derived \cite{Hazeltine1973pp, Hazeltine1978pp} by expressing the velocity, $\Vv$ as combination of a drift and a rotation, as in Appendix~\ref{sec:vspace-avg}, but still using the real space coordinates $\Vx$.
The PDF can be separated into its velocity average and  varying parts
\begin{align}
\pdf =\avgVel{\pdf} + \varyVel{\pdf}
\end{align}
defined in Appendix~\ref{sec:vspace-avg}.
Similarly, the kinetic operator can be decomposed into velocity-average and varying parts as $\CK  =\CK_0+\CKavg_1+\CKvary_1+\dots$  where $\CK_0=\Omgc\partial_\gyrophase$.
The first order gyroaveraged part is
\begin{align}
\CKavg_1 \pdf = \avgVel{\left[\partial_t+ \partial_\| \vpar +\partial_{\mass \vpar}\charge\Epar   \right] }\pdf
\end{align}
and the first order gyrovarying part is
\begin{align}
\CKvary_1 \pdf & =[\CK-\CKavg_1]\pdf\\
&\simeq \left[ \nabla\cdot\Vv_\perp  +\nabla_{\mass\Vv_\perp}\cdot\charge\VE
\right]\pdf
.
\end{align}

The zeroth order equation places a constraint on the gyrovarying part
\begin{align}
\CK_0 \pdf_0=\Omgc \partial_\theta \varyVel{\pdf_\zeroflr }=0
\end{align} 
which implies that the gyrovarying part $\varyVel{\pdf_\zeroflr }$ vanishes.
As explained in Sec.~\ref{sec:why_gyroaverage}, if one begins with the initial guess $\pdf_\zeroflr(\xcoord, \vcoord)$, then, this constraint generates the correction, $\Delta \pdf_\zeroflr =- \varyVel{\pdf_\zeroflr}$, and the full zeroth order solution 
$\avgVel{\pdf_\zeroflr}$.
Higher order FLR corrections to this constraint ultimately generate the true gyroaverage of the initial guess $\avgCycBar{\pdf_\zeroflr}$.

The gyroaveraged first order equation implies that 
\begin{align}
\CKavg_1  \avgVel{\pdf_\zeroflr } =0.
\end{align} 
Often this equation is solved while neglecting the gyroaverage, i.e. in the form $\CKavg_1\pdf_\zeroflr(\xcoord,\vcoord)=0$.
In this case, it generates the initial guess for $\pdf_0$ and then one must return to the zeroth order gyrophase constraint equation above to find the corrected zeroth order solution $\avgVel{\pdf_\zeroflr}$. Higher order FLR corrections ultimately generate the true gyroaverage.

The gyrovarying first order equation implies that 
\begin{align}
\CK_0\varyVel{ \pdf_\firstflr} &=-\CKvary_1 \pdf_\zeroflr\\
\Omgc \partial_\gyrophase\varyVel{ \pdf_\firstflr} &=
-\left[\Vv_\perp  \cdot \Vnabla_\xcoord+ \charge \VE_\perp \cdot \Vnabla_{\mass \vcoord} \right] \pdf_\zeroflr  
\end{align} 
 which has the solution
\begin{align}
\varyVel{ \pdf_\firstflr } =-\CK_0^{-1}\CKvary_1  \pdf_\zeroflr  = - \Vgyro    \cdot\left[\Vnabla+ \charge \VE \, \partial_{\mu \Bfield}\right] \pdf_\zeroflr  .
\end{align} 
Thus, the first order term represents a displacement in position from $\Vx\rightarrow \Vx-\Vgyro=\VR$ and in magnetic moment from $\mu\Bfield\rightarrow \mu\Bfield-\mass\VV_0\cdot\Vv_\perp$ where $\VV_0=\VE\times\bhat/\Bfield$ is the first order drift velocity.
The first order term  leads to both the $\VE\times\VB$ and the first order collective fluid drift velocity since $\Vgyro  =  \Omgc^{-1}\bhat \times   \Vv $.
 
The gyroaveraged second order equation implies that 
\begin{align}
\CKavg_1  \avgVel{\pdf_\firstflr} +{\overline{ \CK}_2} \pdf_0 &=-\avgVel{\CKvary_1\varyVel{\pdf_\firstflr} }
\\
&=    \avgVel{\CKvary_1 \CK_0^{-1} \CKvary_1 }\pdf_\zeroflr
.
\end{align} 
In straight field line geometry, the right hand side vanishes because it is proportional to $(\nabla~\cdot~\bhat~\times~\nabla) \tfrac{1}{2}\gyro^2\pdf_\zeroflr=0$.
Above, $ {\overline{\CK}_2} \pdf_0=\nabla_{\Vx}\cdot\VV_1\pdf_0+\nabla_{\mass\Vv}\cdot\VF_1\pdf_0$ represents the guiding center drift velocity and forces.

 The gyrovarying second order equation implies that 
\begin{multline}
\CK_0 \varyVel{ \pdf_\secondflr } =
-\CKavg_1 \varyVel{\pdf_\firstflr }
-\CKvary_1 \avgVel{\pdf_\firstflr }
-\varyVel{\CKvary_1 \varyVel{ \pdf_\firstflr } }
\\
 =
\CKavg_1 \CK_0^{-1} \CKvary_1 \pdf_\zeroflr 
-\CKvary_1 \avgVel{\pdf_\firstflr }
\\
+\varyVel{\CKvary_1 \CK_0^{-1} \CKvary_1 \pdf_\zeroflr  }
.
\end{multline} 
To clarify notation in the following, let us write the solution as $\varyVel{ \pdf_{\secondflr }}=\varyVel{ \pdf_{\secondflr,0}} +\varyVel{ \pdf_{\secondflr,1}}+\varyVel{ \pdf_{\secondflr,2}}$ where
\begin{multline}
\varyVel{ \pdf_{\secondflr,0}}   =- \CK_0^{-1} \CKavg_1 \varyVel{\pdf_\firstflr } =\CKavg_1 \CK_0^{-2} \CKvary_1 \pdf_\zeroflr 
\\
 =- \CKavg_1 \Omgc^{-2} \left[\Vnabla_\xcoord \cdot  \Vv_\perp  +  \nabla_{\mass\vcoord} \cdot \charge \VE_\perp \right] \pdf_\zeroflr 
,
\end{multline} 
\begin{multline}
 \varyVel{ \pdf_{\secondflr,1}}   =-\CK_0^{-1} \CKvary_1 \avgVel{\pdf_\firstflr } 
\\
 -\left[\Vnabla_\xcoord \cdot  \Vgyro+  \nabla_{\mass\vcoord} \cdot \charge \VE_\perp \right]     \avgVel{\pdf_\firstflr } 
 ,
\end{multline}  
and
\begin{multline}
 \varyVel{ \pdf_{\secondflr,2}}  =
 -\varyVel{\CK_0^{-1}\CKvary_1 \varyVel{ \pdf_\firstflr } }
 =  \varyVel{\CK_0^{-1} \CKvary_1 \CK_0^{-1} \CKvary_1 \pdf_\zeroflr  }
\\
= -\left[\Vnabla_\xcoord +  \partial_{\mu \Bfield} \charge \VE \right]\cdot
  \left[ \Vgyro \Vgyro-\tfrac{1}{2}\gyro^2\unitTensor_\perp \right]\cdot
 \\
\left[\Vnabla_\xcoord+  \charge \VE\, \partial_{\mu \Bfield}  \right]  \pdf_\zeroflr
 .
 \end{multline} 
 Where, in the final expression, we use the fact that, assuming that $\pdf_0$ only depends on $\mu\Bfield=\mass\Vv_\perp^2/2$ rather than $\Vv_\perp$, allows one to express the perpendicular velocity gradient   as
\begin{align}
\Vnabla_{\mass \vcoord}\cdot \VE_\perp \pdfstart=\VE_\perp \cdot \Vv_\perp \, \partial_{\mu \Bfield}\pdfstart .
\end{align}

The third order gyroaveraged equation is
\begin{multline}
\CKavg_1\avgVel{\pdf_2}+\CKavg_2\avgVel{\pdf_1}+\CKavg_3\avgVel{\pdf_0}=
\\
 -\avgVel{\CKvary_1\varyVel{\pdf_2}} - \avgVel{\CKvary_2\varyVel{\pdf_1}}.
\end{multline}
The first term on the right hand side generates polarization through the $\varyVel{ \pdf_{\secondflr,0}}$ term in $\varyVel{ \pdf_{\secondflr}}$ because this is the part that is operated on by $\CKavg_1$.
The last term on the left hand side $\CKavg_3\pdf_0:=\nabla_{\Vx}\cdot\VV_2+\nabla_{\mass\Vv}\cdot\VF_2$ requires a computation of the second order drifts including higher derivative and nonlinear terms.
The last term on the right requires a careful calculation of the varying part of the guiding center drift.
These calculations are more easily performed using gyrokinetic theory, to be discussed in Sec.~\ref{sec:gk}.

The polarization itself is relatively simple to calculate. In phase space, it has the form  
\begin{align}
\CKavg_1 \pdf_{\secondflr, \pol} &:= -\CKvary_1 \varyVel{ \pdf_{\secondflr,0}} 
\\
  &=-\left[\Vnabla_\xcoord \cdot \Vv_\perp +\Vnabla_{\mass \vcoord}  \cdot \charge \VE_\perp \right] \varyVel{ \pdf_{\secondflr,0}}  .
\end{align}
The solution to this equation can be found by integrating in time to find
\begin{multline}	
\pdf_{\secondflr, \pol} = \CKvary_1\CK_0^{-2}\CKvary_1\pdf_0
\\
=\Vnabla_\xcoord \cdot  \Omgc^{-2} \Vv_\perp \left ( \Vv_\perp  \cdot \Vnabla_\xcoord  +\charge \VE_\perp \cdot\Vnabla_{\mass \vcoord}\right)\pdfstart 
\\
+\Vnabla_{\mass \vcoord}\cdot \charge \VEfield_\perp  \Omgc^{-2}   \left(\Vv_\perp  \cdot \Vnabla_\xcoord +\charge \VE_\perp\cdot \Vnabla_{\mass \vcoord}\right)\pdfstart.
\end{multline}
 The gyroaveraged part of this expression, which is needed to calculate the density in real space, 
 reduces to 
\begin{multline}
\avgVel{\pdf_{\secondflr, \pol}}  
=\Vnabla_\xcoord \cdot \tfrac{1}{2} \gyro_\perp^2   \left(\Vnabla_\perp +\charge \VEperp\partial_{\mu \Bfield}\right)\pdfstart
\\
+ \partial_{\mu \Bfield} \charge \VEfield_\perp \cdot  \tfrac{1}{2}\gyro_\perp^2   \left(\Vnabla_\perp +\charge \VEperp\partial_{\mu \Bfield}\right)\pdfstart
.
\end{multline}
This is simply the result of the  second order displacement in position  $\Vx\rightarrow \Vx-\Vgyro$ and magnetic moment $\mu\Bfield\rightarrow \mu\Bfield -\mass\Vv_\perp\cdot\VV_0$.

Integration over velocity space yields the second order polarization density
 \begin{align}
\density_{\secondflr, \pol } &=\int  \pdf_{\secondflr, \pol}  d^3v=\int \avgVel{\pdf_{\secondflr, \pol}} d^3v\\
&=\Vnabla\cdot     (\Vnabla_\perp \press_{\zeroflr \perp}-\charge\densitystart  \VEperp )\, \,  \gyro_T^2/\Tperp .
\end{align}
This establishes the validity of Eq.~\ref{eq:pol_drift} using the drift kinetic FLR expansion.

 If we had started with the assumption $\pdf_0(\VR,\Vw)$, this would have simplified the results.  However, it would not have generated the gyroaveraged contribution shown above.  Rather, it would yield the same second-order result as the GC/GK picture, discussed next.

 \section{Gyrokinetic Theory \label{sec:gk}}

\subsection{Action-Angle Coordinates \label{sec:gk_trans}} 

The goal of this section is to calculate the polarization using adiabatic  theory in gyrokinetic (GK) action-angle coordinates.
The GK polarization charge density is derived in the next subsection.
The polarization of linear perturbations to the equilibrium is defined and calculated in Sec.~\ref{sec:gk_linear}.
This allows one to understand the polarization of waves and instabilities on solid footing.
Applications are discussed in Sec.~\ref{sec:applications}, including the treatment of anisotropic Maxwell-Boltzmann distribution functions.

The calculation  begins and ends in real space coordinates, $\zcoord=\{x,v\}$.
The first step is to transform to the lowest order guiding center coordinates, $\Zcoord :=\CT(\zcoord)= \{\Rcoord,\Vpar,\Jcyc,\gyrophase\}$, defined in Appendix~\ref{sec:gc_coordinates}.
The action of the transformation on differential forms is defined via  the pullback $\pdf(\zcoord)=\ST\PDF$ and the pushforward $\PDF(\Zcoord)=\ST^{-1} \pdf$ discussed in Appendix~\ref{sec:gk-transformation}. 

The next step \cite{Dubin83pof} is to solve for a transformation of the guiding center coordinates $\Zcoord$ to a new set of coordinates $\GKcoord = \{\Rbar,\bar \Vpar,\Jbar,\bar \gyrophase\}$ that ensure that that the new magnetic moment $\Jbar$ remains a true adiabatic invariant, order by order in $\epsilon$. 
The transformation to gyrokinetic coordinates is described in more detail in the appendices.
Again, in terms of the action, the magnetic moment is $\mu=\charge\Jcyc/\mass$.
We also simplify the discussion here by assuming that the magnetic field $\VB$ is constant.

In order to focus on the long timescale dynamics, for frequencies well below the gyrofrequency,  the GK PDF,  $\bar\PDF(\bar\Zcoord)$, can be assumed to be independent of $\bar \theta$  in the new coordinate system,  so that it does not have any fast  temporal behavior \cite{Dubin83pof}.  
The PDF in any other coordinate system, e.g. $\pdf(\zcoord)$ or $\PDF(\Zcoord)$, will appear to have a very specific gyro-phase dependence that is intrinsically locked to the gyroaveraged distribution by the oscillatory part of the particle motion.  
The gyrophase dependence in real space leads to the finite orbit width effects of polarization and magnetization.

The  two largest terms in the kinetic equation yield
\begin{align}
\Omgc \partial_{\gyrophase} \PDF -  \partial_{\gyrophase} \charge\phi \partial_{\Jcyc} \PDF  =0
\end{align}
which can be solved by the method of characteristics. This leads to the simple transformation 
\begin{align}
 \Jbar=\Jcyc+\varyCyc{\charge \phi}/\Omgc= \mass \vperp^2/2 + \varyCyc{\charge \phi}/\Omgc
\end{align} 
  where the gyro-varying part of the potential is
\begin{align}
\varyCyc{ \phi} = \phi(\xcoord)-\avgCyc{\phi},
\end{align}  
i.e., it is the potential without its gyroaveraged part $\avgCyc{\phi}$ (defined in Appendix~\ref{sec:gyroaverage}).  In a Hamiltonian treatment of gyrokinetics, one will also redefine the gyrophase in order to keep the phase space coordinates canonical.  This can be performed if one defines the mixed variable generating function, $S$,  for the canonical transformation  \cite{Lichtenberg92book} to be 
\begin{align}
\Action=\Jbar \gyrophase  - \int \varyCyc{\charge \phi}  d\gyrophase/\Omgc.
\end{align}
In this case, the two coordinates are determined via
\begin{align}
 \Jcyc &= \partial_\gyrophase S = \Jbar - \varyCyc{\charge \phi} /\Omgc \\
 \bar \gyrophase&= \partial_{\Jbar}  S = \gyrophase - \int \partial_\Jbar \varyCyc{\charge \phi} d\gyrophase/\Omgc .
\end{align} 
The generating function leads to the coordinate system $\GKcoord = \{\bar \Rcoord, \bar \Vpar,\Jbar, \bar\gyrophase\}$.

In this section, the amplitude expansion  will be denoted via
\begin{align}
\PDFbar =\PDFbar_\eq +\overline{ \PDFpert }+  \dots
\end{align}
where $\PDFbar_\eq(\bar \Rcoord,\bar \Vpar,\bar \Jcyc)$ is the equilibrium PDF in GK coordinates and $\pdf_\eq(\xcoord,\vcoord)=\PDFbar_\eq(\GKcoord(\xcoord,\vcoord))$ is the equilibrium PDF in real space. 
 In GC coordinates, the equilibrium PDF itself has the linear response
\begin{align}
 \PDF_\eq &= \PDFbar_\eq - \varyCyc{\charge\phipert} \partial_{\Jcyc \Omgc } \PDFbar_\eq +  \dots
\end{align}
so that the total linear perturbation is
\begin{align}
 \PDFpert &= \overline{\PDFpert}  - \varyCyc{\charge\phipert}  \partial_{\Jcyc \Omgc } \PDFbar_\eq
 +\dots
 .
\end{align}
Finally, one transforms back to real space via $\pdfpert=\PDFpert(\Zcoord(\xcoord,\vcoord))$. The real space particle density is given by the real space gyroaverage
\begin{align}
\densitypert &=  \int   \avgVel{ \pdfpert} d^3v = \int  \avgVel{  \PDFpert}  \delta^3(\Vx -\VR -\Vgyro)d^3xd^3v.
\end{align}

\subsection{ Gyrokinetic Charge Density \label{sec:gk_noether} }

 The charge density is determined by the application of N\"oether's theorem to the Lagrangian of the theory.
 If the action is defined via the Lagrangian density $\Action=\int\Lagrangian d^3\xcoord$, then 
 $\chargedensity:=-\delta \Lagrangian/\delta \phi $.
 
 In real space, the charge density is precisely determined by the electrostatic potential energy term $ \phipert  \charge \int d^3\vcoord \pdf$ which yields the usual definition.
 In action-angle coordinates, this same term becomes more complicated
\begin{align}
\int d^3\xcoord \int d^3\vcoord  \phipert \avgVel{ \ST \PDF  }=
\int d^3\xcoord \int d^3\vcoord   \charge \phipert \SJ_0   \ST  \PDF 
\end{align}
where $\SJ_0$ is the gyroaveraging operator defined in Appendix~\ref{sec:gyroaverage}.
Thus, this identifies the density of charge carriers as 
\begin{align}
\chargedensity=\charge\density =\charge\int d^3\vcoord  \avgVel{\ST\PDF} = \charge\int d^3\vcoord \SJ_0 \ST\PDF.
\end{align}

\subsection{Linear Gyrokinetic Polarization \label{sec:gk_linear} }

 In gyrokinetic action-angle coordinates,  the fast evolution of the gyrophase separates from the slow evolution of the gyroaveraged distribution function.  
 The  collisionless  GK equation is
\begin{align}
\left[\partial_t +\Vnabla_{\bar \Rcoord}\cdot  \left(\vpar \bhat+\avgCycBar{\VV_E}\right) + \partial_{\mass\vpar}  \avgCycBar{\charge\Efield_{\|} } \right]   \PDFbar=0
\end{align}
where the electric drift velocity is 
$\VV_E~=  \VE~\times~\bhat~/~\Bfield$.
The linearized GK equation is
\begin{multline}
\left(\partial_t +\Vnabla_{ \Rcoord}\cdot   \vpar \bhat  \right)    \overline{\PDFpert} =\\
-\left( \Vnabla_{ \bar \Rcoord}\cdot  \avgCycBar{\pert \VV_E}  + \partial_{\mass\vpar} \avgCycBar{\charge\pert \Efield_{\|}} \right) \PDFbar_\eq.
\end{multline}
Because the gyroaveraging operation  commutes with all  terms in the linearized GK equation, the solution can be written as 
\begin{align}
\overline{\PDFpert} =\avgCycBar{\STzero^{-1} \pert \gfun_0} \simeq \avgCyc{\STzero^{-1} \pert \gfun_0},
\end{align}
 where the PDF in  the limit of ``zero orbit width,''  $\pert \gfun_0$, satisfies the linear evolution equation 
\begin{align}
\left(\partial_t +\Vnabla_ \xcoord \cdot   \vpar \bhat  \right)     {\pert\gfun}_\zeroflr =
-\left( \Vnabla_{\xcoord}\cdot  \VVpert_E  + \partial_{\mass\vpar} \charge \Epert_{\|} \right)     \PDF_\eq .
\end{align}
Since $\lim_{\gyro \rightarrow 0} \Rcoord = \xcoord$,   this equation can be considered to be formulated in real space.   

Thus, the total zeroth order perturbation can be defined as
\begin{align}
\pert \pdf_\zeroflr ={\pert\gfun}_\zeroflr - \charge\phipert\partial_{\Jcyc \Omgc} \PDF_\eq.
\end{align}
and  the effect of polarization on the PDF to first order in $\delta$  can be defined as
\begin{align}
\pdfpert_{\pol }   =\ST \avgCyc{\STzero^{-1} {\pdfpert}_\zeroflr} -  \pdfpert_\zeroflr =-\ST \varyCyc{\STzero^{-1} \pdfpert_\zeroflr} .
\end{align}

Due to the small amplitude ($\delta$) expansion, the average can be separated into an electric and a \kinetic contribution
\begin{align}
\pdfpert_{\pol } =\pdfpert_{\pol,\Efield } +\pdfpert_{\pol,\kin } .
\end{align}
Using  Eq.~\ref{eq:gyroavg_first}, the electric part is
\begin{align}
\pdfpert_{\pol,\Efield }  &=   -  \ST \avgCyc{\STzero^{-1} \charge \phipert}  \ST \partial_{\Jcyc \Omgc} \PDF_\eq + \charge \phipert  \partial_{\Jcyc \Omgc}\pdf_\eq\\
&=\ST \varyCyc{\STzero^{-1} \charge \phipert} \ST \partial_{\Jcyc \Omgc} \PDF_\eq
\end{align}
while the \kinetic polarization is
\begin{align}
\pdfpert_{\pol,\kin }  =\ST \avgCyc{ \STzero^{-1} {\pert \gfun}_\zeroflr} - \ST\STzero^{-1}{\pert \gfun}_\zeroflr =-\ST \varyCyc{ \STzero^{-1} {\pert \gfun}_\zeroflr}.
\end{align}
Note that if one were to try to identify the \kinetic polarization as $\avgVel{\ST \PDFpert}-\PDFpert=-\varyVel{\PDFpert}$, then one would only obtain 1/2 the magnitude of the correct result. 

The velocity angle average of the solution in real space, $\pdfpert$, 
\begin{align}
\avgVel{\pdfpert}  = \avgVel{\ST \PDFpert} = \avgVel{\ST \avgCyc{ \STzero^{-1}  {\pert \pdf}_\zeroflr}} =\SJ_0\ST\SJ_0\STzero^{-1} {\pert \pdf}_\zeroflr 
\end{align}
determines the density in real space 
\begin{align}
\densitypert =\int \avgVel{\ST \avgCyc{\STzero^{-1} {\pert \pdf}_\zeroflr}} d^3v =\int \SJ_0\ST\SJ_0\STzero^{-1} {\pert \pdf}_\zeroflr d^3v.
\end{align}
The polarization density is defined by the relation
\begin{align}
\densitypert=\densitypert_0+\densitypert_\pol
\end{align}
where
\begin{align}
 \densitystartpert (\xcoord)=\int  {\pert \pdf}_\zeroflr(\xcoord,\vcoord)d^3v.
\end{align}
Hence, the polarization density is
\begin{align}
\densitypert_{\pol }  &= \int \left[\avgVel{\ST \avgCyc{ \STzero^{-1} {\pdfpert}_\zeroflr}} - \ST \STzero^{-1} \pdfpert_\zeroflr\right] d^3v\\
&=- \int  \avgVel{ \ST \varyCyc{\STzero^{-1} \pdfpert_\zeroflr}}   d^3v .
\end{align}
To first order in $\delta$, one can separate the polarization density
\begin{align}
\densitypert_\pol = \densitypert_{\pol,\Efield } +\densitypert_{\pol ,\kin} 
\end{align}
 into electric and \kinetic contributions, where 
\begin{align}
   \densitypert_{\pol,\Efield} &=  \int  \avgVel{\pdfpert_{\pol,\Efield}} d^3v \\
   \densitypert_{\pol,\kin} &=  \int  \avgVel{\pdfpert_{\pol,\kin}} d^3v .
\end{align}
Thus, this yields the polarization densities
 \begin{align}
\densitypert_{\pol,\Efield } 
&= \int  \avgVel{ \ST \varyCyc{\STzero^{-1} \charge \phipert}  \partial_{\Jcyc \Omgc} \PDF_\eq}  d^3v
\\
\densitypert_{\pol,\kin}  
&= -  \int  \avgVel{ \ST \varyCyc{ \STzero^{-1} {\pert \gfun}_\zeroflr} } d^3v
.
\end{align}

To second order in $\kperp \gyro$, the effect on the PDF is
\begin{align}
\PDFpert&=\avgCyc{\STzero^{-1}\pdfstartpert} -\pdfstartpert    \simeq  \tfrac{1}{4} \Vnabla_\perp\cdot  \gyro^2 \Vnabla_\perp \pdfstartpert  \\
\avgVel{\pdfpert}&= \avgVel{\ST \avgCyc{\STzero^{-1}\pdfstartpert}}-\pdfstartpert    \simeq    \tfrac{1}{2} \Vnabla_\perp\cdot  \gyro^2 \Vnabla_\perp \pdfstartpert 
\end{align}
and, hence,
\begin{align}
\avgVel{\pdfpert_{\pol,\Efield}} & \simeq   \tfrac{1}{2} \Vnabla_\perp\cdot  \gyro^2  \PDF_\eq \Vnabla_\perp\charge \phipert/\Tperp \\
\avgVel{\pdfpert_{\pol,\kin}} & \simeq   \tfrac{1}{2} \Vnabla_\perp\cdot  \gyro^2 \Vnabla_\perp \pdfstartpert .
\end{align}
This implies that the electric polarization density is
\begin{align}
\densitypert_{\pol,\Efield}&= \Vnabla \cdot \density_0 \gyro_T^2\left( \ \Vnabla_\perp \charge \phipert_\perp \right)/\Tperp
.
\end{align}
To second order in $\kperp \gyro$,  the \kinetic polarization is entirely due to the diamagnetic polarization density, defined via
\begin{align}
\densitypert_{\pol,\kin}&= \Vnabla \cdot \gyro_T^2\left( \Vnabla_\perp \presspert_{0\perp}\right) /\Tperp
=:\densitypert_{\pol,\diamag}.
\end{align}
Thus, this leads to Eq.~\ref{eq:pol_drift} and Eq.~\ref{eq:pol_gyro_corrected}.

The guiding center density is defined as
\begin{align}
 \Densitypert(\Rcoord)&= \int  \PDFpert d\vpar d\Jcyc \Omgc d\gyrophase =\int \avgCyc{\STzero^{-1} \pdfpert_\zeroflr} d\vpar d\Jcyc \Omgc d\gyrophase .
\end{align}
If one defines the polarization density by the expression 
\begin{align} \label{eq:gc_pol_full}
\Densitypert_{\pol}(\xcoord) &=\densitypert(\xcoord) -\Densitypert (\xcoord) 
\end{align}
then, to second order in FLR, the  electric and diamagnetic terms are  
\begin{align}
\Densitypert_{\pol,\Efield} (\xcoord)&= \Vnabla \cdot  \gyro_T^2  \Density \left(  \Vnabla_\perp \charge \phipert \right)/\Tperp \\
\Densitypert_{\pol,\kin} (\xcoord)&= \tfrac{1}{2} \Vnabla \cdot  \gyro_T^2\left(  \Vnabla_\perp \Presspert_\perp\right)\Tperp 
=:\Densitypert_{\pol,\diamag} (\xcoord)
\end{align}
which establishes the validity of Eq.~\ref{eq:pol_gyro}.
However, the real space polarization density, $ \densitypert_{\pol}=\Densitypert_\pol+\overline{ \Densitypert}_{\pol,\kin}$, also includes the contribution
\begin{align}\label{eq:gk_pol_full}
 \overline{\Densitypert}_{\pol,\kin}(\xcoord) &= \Densitypert(\Rcoord)- \densitypert_\zeroflr(\xcoord)  
 \\
 &=   \tfrac{1}{2} \Vnabla \cdot \gyro_T^2\left( \Vnabla_\perp \Presspert_\perp \right) /\Tperp =: \overline{\Densitypert}_{\pol,\diamag}
 .
\end{align}
Since, at this order, $ \overline{\Densitypert}_{\pol,\kin}= \Densitypert_{\pol,\kin}=\densitypert_{\pol,\kin}/2$, the full real space polarization density is actually given by  Eq.~\ref{eq:pol_gyro_corrected}.


 \subsection{Gyrokinetic Applications \label{sec:applications}}

 \subsubsection{Anisotropic Maxwellian \label{sec:gk_maxwellian}}
 
Consider basing the PDF on an anisotropic Maxwellian equilibrium distribution function defined via
\begin{align}
\pdf_M =   e^{-\mass\vpar^2/2\Tpar - \Jcyc \Omgc/\Tperp}/(2\pi\Tpar)^{1/2}2\pi\Tperp
\end{align}
and  $\Tpar$ and $\Tperp$ are constant in space.
The zeroth order PDF will be assumed to have the form
\begin{align}
\pdfstart  =\densitystart (\xcoord) \pdf_M.
\end{align}
For constant $\VB$ and constant perpendicular temperature $\Tperp$, one can include all orders in $\kperp\gyro$ analytically \cite{Stix92book}. 
In this case, $\Vpar=\vpar$ and $\Jcyc=\charge\mu/\mass=\mass\vperp^2/2\Omgc$.
Averaging over the gyrophase yields the expression
\begin{align}
\PDF(\Rcoord, \Vpar, \Jcyc) &= \oint \densitystart(\xcoord) \pdf_M e^{-i\Vk \cdot(\Vx-\VR) } \frac{d^3\xcoord  d\gyrophase}{(2\pi)^{5/2} }
\\
&= \int \pdf_M 
 J_0(b)\densitystart(\kcoord)  e^{i\Vk\cdot\VR} d^3k/(2\pi)^{3/2}
\end{align}
where $J_0(b)$ is the Bessel function of order 0, 
\begin{align}
b^2:=(\Vkperp\cdot\Vgyro)^2=-\gyro^2\nabla_\perp^2,
\end{align}
and  
\begin{align}
\nabla_\perp^2:=\nabla\cdot(\unitTensor-\bhat\bhat)\cdot\nabla.
\end{align}
We can also consider $J_0$ to denote the linear operator defined by the Fourier integral above.
This allows us to state that
\begin{align}
\PDF =\pdf_M  J_0(b)\densitystart .
\end{align}
Integration over the Maxwellian in velocity space yields the gyrofluid density \begin{align}\label{eq:gf_density}
\Density(\Rcoord)&=\left. e^{-b_T^2/2}   \densitystart \right|_{\Rcoord}
\end{align}
where
 \begin{align}
 b_T^2=\gyro_T^2\kperp^2=-\gyro_T^2\nabla_\perp^2
\end{align}
and $\gyro_T=\Tperp/\mass \Omgc^2$ is the thermal gyroradius. 
The PDF in GK coordinates, $\PDF$,  is  the gyroaverage of $\densitystart$ so that
\begin{align}
\PDF =\avgCyc{\pdf_0} =\pdf_M J_0(b)\densitystart =\pdf_M J_0(b)e^{b_T^2/2}\Density.
\end{align}

As explained in Appendix~\ref{sec:gk-transformation}, the density in real space is defined by
\begin{align}
\pdf(\xcoord,\vcoord) &= \oint \PDF(\Rcoord,\Vpar,\Jcyc) e^{i\Vk \cdot(\Vx-\Vgyro) }  d^3\Rcoord /(2\pi)^{3/2}
\\
&= \sum_n \int J_{-n}(b) \PDF(\kcoord, \Vpar, \Jcyc) e^{i\Vk\cdot\Vx + in(\gyrophase-\gyrophase_\kcoord)} \frac{d^3\kcoord}{(2\pi)^{3/2}}
\end{align}
where $\tan{(\gyrophase_\kcoord)}:=-\Vk\cdot\ahat/\Vk\cdot\chat$.
Hence, the gyroaveraged real space PDF is 
\begin{align}
\avgVel{\pdf} =  J_0 \PDF = \pdf_M J_0^2 \densitystart.
\end{align}
Integrating over the Maxwellian PDF yields the real space density
 \begin{align}
 \density(\xcoord) =\Gamma_0(b_T^2) \densitystart =\Gamma_0(b_T^2) e^{b_T^2/2} \Density.
 \end{align}
Here, $\Gamma_0$ is defined via 
 \begin{align}
 \Gamma_0(b_T^2)=e^{-b_T^2}I_0(b_T^2)
 \end{align}
 where $I_0$ is the modified Bessel function of order 0.
 It is also considered to be a linear operator through it's definition as a Fourier transform.
 
The factor of 1/2 difference  in polarization arises from the question of whether $\density$ is being compared to $\Density$ or to $\density_0$. 
The gyroaveraged  $\avgCyc{\pdf}$ is related to $\PDF$ by the factor $J_0(b)\simeq 1-b^2/4  $ while it is related to $\pdfstart$ by the factor  $J_0^2(b)\simeq 1-b^2/2$.
Equivalently, $\density$ is related to $\Density$ by integrating $J_0(b)$ over a Maxwellian, which yields the factor $e^{-b_T^2/2}\simeq 1-b_T^2/2$.
In constrast, comparing $\density$ to $\density_0$ results from integrating $J_0^2(b)$ over a Maxwellian, which yields the factor $\Gamma_0(b_T^2)=e^{-b_T^2} I_0(b_T^2) \simeq 1-b_T^2$. 
  
Assume that the density can be expanded to first order in amplitude as $\density = \density_{\eq}+\densitypert$, where $ \density_{\eq}$ is constant.
In this case,  the linear electric polarization density is given by
\begin{align}
\densitypert_{\pol,\Efield} &= (\Gamma_0-1 )\density_\eq \charge\phipert/\Tperp.
\end{align}
The perturbed density in real space is related to the zeroth order density via
\begin{align}
\densitypert =  \Gamma_0 {\densitypert}_\zeroflr 
\end{align}
and, hence, the \kinetic polarization density is  
\begin{align}
\densitypert_{\pol,\kin} = ( \Gamma_0-1){\pert \density}_\zeroflr .
\end{align}
Thus, the linear response relation is 
\begin{align}
\densitypert = \Gamma_0 {\pert \density}_\zeroflr+  (\Gamma_0-1 ) \density_{\eq} \charge\phipert/\Tperp
\end{align}
and the total polarization density is
\begin{align}
\densitypert_\pol = (\Gamma_0-1) {\pert \density}_\zeroflr +(\Gamma_0-1 ) \density_{\eq} \charge\phipert/\Tperp.
\end{align}

If one were to rewrite the result  for the  real space \kinetic polarization density in terms of the guiding center (GF) density, then one obtains
 \begin{align}
\densitypert_{\pol,\kin} &= ( \Gamma_0-1) e^{b_T^2/2}\Densitypert.
\end{align}
This agrees with Eq.~\ref{eq:pol_gyro} to second order in $\kperp\gyro$. 
Note, however, that this expression only converges for sufficiently smooth $\Densitypert$, such as that implied by Eq.~\ref{eq:gf_density}.
Because the gyroaverage is a lossy operation, it is a more stable procedure to first define $\density_\eq$, $\densitypert_0$, and $ \densitypert$ and to then derive $\Densitypert$.

If there is no temperature anisotropy, then the net polarization $\densitypert_\pol$ must vanish.
In the presence of a temperature anisotropy,
  the Boltzmann relation  $\densitypert_\zeroflr =-\density_{\eq} \charge \phipert/\Tpar$ 
yields
\begin{align}
\densitypert_\pol &= (\Gamma_0-1 ) \density_{\eq} \charge\phipert\times(1/\Tperp-1/\Tpar)\\
&= (\Gamma_0-1 ) \density_{\eq} \charge\phipert \times (\Delta \Temp/\Tpar\Tperp).
\end{align}
Thus, the net polarization is proportional to the temperature anisotropy $\Delta\Temp=\Tpar-\Tperp$.

Due to the nonlinearity, one cannot derive closed-form results when there is a temperature gradient.
In fact, for general flows, the same is true for velocity gradients.
If there is a spatial variation in temperature, then, the result to second order in $\kperp\gyro$ is
\begin{align}
  \densitypert_{\pol,\kin} = \nabla_\perp^2\press_\perp/\mass\Omgc^2.
\end{align}

\subsubsection{Anisotropic Maxwell-Boltzmann \label{sec:maxwell-boltzmann}}

Now, consider the anisotropic Maxwell-Boltzmann distribution
\begin{align}
\pdf_{MB}  (\xcoord,\vcoord) &=  \frac{e^{-\Ham/\Tpar+\Jcyc\Omgc(1/\Tpar-1/\Tperp)} }{(2\pi \Tpar)^{1/2}
2\pi\Tperp}
\\
&=  \frac{e^{-\Ham/\Tpar+\Jcyc\Omgc \Delta\Temp/\Tpar\Tperp} }{(2\pi \Tpar)^{1/2}2\pi\Tperp}.
\end{align}
This can also be written in terms of the anisotropic Maxwellian as well as in terms of the  the PDF in the limit of vanishing gyroradius
\begin{align}
\pdf_\zeroflr=e^{-\charge \phi/\Tpar}
\end{align}
via
\begin{align}
\pdf_{MB}
&=e^{-\avgCyc{\charge \phi}/\Tpar - \varyCyc{\charge \phi}/\Tperp}\pdf_M
\\
&=e^{ \varyCyc{\charge \phi}/\Tpar - \varyCyc{\charge \phi}/\Tperp}\pdf_\zeroflr.
\end{align}

 This is defined in terms of the Hamiltonian, which is assumed to vary sufficiently slowly in time that it can be considered to be conserved.
In this case, the Hamiltonian is unmodified by the gyroaverage 
\begin{multline}
\Ham=\avgCyc{\Ham} = \tfrac{1}{2}\mass \vel^2 + \charge\phi 
= \tfrac{1}{2} \mass \vpar^2 + \mu \Bfield + \charge\phi 
\\
=\tfrac{1}{2} \mass \vpar^2 +\Jcyc \Omgc +  \avgCyc{\charge\phi}
\end{multline}
and, hence, for constant $\Temp$, the Maxwell-Boltzmann distribution is also unmodified by the gyroaverage
\begin{align}
 \pdf_{MB}=\avgCyc{\ST^{-1} \pdf_{MB}} .
\end{align}

Let the zeroth order distribution function be based on the Maxwell-Boltzmann distribution
\begin{align}
\pdfstart(\xcoord,\vcoord) =\pdf_{MB}(\xcoord,\vcoord) \eta_0 (\xcoord)
\end{align}
so that
\begin{align}
\densitystart = e^{-\charge\phi/\Tpar} \eta_0 (\xcoord).
\end{align}
Then, following the same line of reasoning as in the previous subsection, one finds that the PDF in action-angle coordinates is
\begin{align}
\PDF(\Rcoord, \Vpar,\Jcyc) &= J_0(b) \pdfstart =\pdf_{MB} J_0\eta_0 \\
&= \pdf_{M} e^{-\charge\phi/\Tpar} J_0 \left[e^{\charge\phi/\Tpar} \densitystart\right].
\end{align}

Now the Boltzmann factor for the gyrofluid density depends on $\charge \phi(\Rcoord)$ if cylindrical $\{\mu,\gyrophase\}$ coordinates are used, but  depends on $\avgCyc{\charge\phi}$ if the more accurate gyrokinetic $\{\Jbar,\bar\gyrophase\}$ coordinates are used. 
While the latter choice may be more accurate, the resulting expressions will depend on the potential itself and cannot be determined in closed form.
If cylindrical $\{\mu,\gyrophase\}$ coordinates are used, then the gyrofluid density has the simple closed-form expression
\begin{align}
\Density(\Rcoord) &=\left. e^{-\charge\phi/\Tpar} e^{-b_T^2/2}\eta_0 \right|_\Rcoord
\\
&=\left. e^{-\charge\phi/\Tpar} e^{-b_T^2/2} e^{ \charge\phi/\Tpar} \density_0 \right|_\Rcoord
\end{align}
which can be inverted to find
\begin{align}
\eta_0(\xcoord) &=\left. e^{b_T^2/2} e^{ \charge\phi/\Tpar} \Density \right|_\xcoord
\\
\density_0(\xcoord) &=\left. e^{ -\charge\phi/\Tpar} e^{b_T^2/2} e^{ \charge\phi/\Tpar} \Density \right|_\xcoord
.
\end{align}

As explained in Appendix~\ref{sec:gk-transformation}, the PDF in real space is
\begin{align}
\pdf (\xcoord,\vcoord)&= \sum_{n}  e^{in(\gyrophase-\gyrophase_\kcoord)} J_{-n}(b) \PDF
 \\
 &=\pdf_{MB} \sum_{n}  e^{in(\gyrophase-\gyrophase_\kcoord)} J_{-n} J_0  e^{b_T^2/2} e^{\charge\phi/\Tpar} \Density
 \\
& =\pdf_{MB} \sum_{n}  e^{in(\gyrophase-\gyrophase_\kcoord)}  J_{-n} J_0\eta_0 \\
&= \pdf_{M} \sum_{n}  e^{in(\gyrophase-\gyrophase_\kcoord)-\charge\phi/\Tpar}  J_{-n} J_0\left[e^{ \charge\phi/\Tpar}\densitystart \right] 
.
\end{align}
The gyroaveraged part is
\begin{align}
\avgCyc{\pdf }(\xcoord,\vcoord)& =   J_0(b) \PDF  =\pdf_{MB} J_0^2 e^{b_T^2/2} e^{\charge\phi/\Tpar} \Density 
\\
&=\pdf_{MB} J_0^2 \eta_0 
\\
&= \pdf_M e^{-\charge\phi/\Tpar} J_0^2\left[e^{-\charge\phi/\Tpar}\densitystart \right] 
.
\end{align}
Finally, the real space density is given by
\begin{align}
\density(\xcoord) 
&=\left. e^{-\charge\phi/\Tpar}   \Gamma_0(b_T^2)    e^{b_T^2/2}\left[ e^{\charge\phi/\Tpar} \Density\right] \right|_\xcoord
\\
&=e^{-\charge\phi/\Tpar}   \Gamma_0  \eta_0
\\
&= e^{-\charge\phi/\Tpar}   \Gamma_0 \left[e^{\charge\phi/\Tpar} \density_0\right]
.
\end{align}
Thus, the exact nonlinear net polarization density is
\begin{align}
\density_\pol =\density -\density_\zeroflr
&=\left[ e^{-\charge\phi/\Tpar}   \Gamma_0 e^{\charge\phi/\Tpar} -1\right] \density_0
.
\end{align}
This agrees with the results of the previous subsection 
to first order in amplitude.

\section{General Kinetic Analysis  \label{sec:kinetic} }

\subsection{Kinetic Equilibrium  \label{sec:kinetic_equilibrium} }

In the absence of strong sources and sinks, a collisionless equilibrium can only have strong dependence on exactly conserved constants of the motion and well-conserved adiabatic invariants.
In the following, we will estimate the size of the small corrections to this statement that are driven by sources and nonlinearities.

However, because collisions are highly localized in space and time, they only preserve constants of the motion, i.e. invariants, that are also local in space and time.
This rules out a strong dependence of the equilibrium on nonlocal adiabatic invariants such as the magnetic moment or the parallel bounce invariant that \emph{require} averaging over a trajectory for their definition, i.e. nonlocal invariants that \emph{require} the integral form $\Jcoord=\oint \Vp\cdot d\Vx/2\pi$ where $\Vp$ is the canonical momentum.

Thus, the kinetic equilibrium for any spatial region that has good particle confinement, i.e. that is confined for times longer than the collision time, can only have strong dependence on \emph{local constants of the motion}.
The presence of gradients in the PDF will add kinetic corrections to this results, but, for well-confined plasmas, these additional terms must be small at both high and low collisionality.

\subsection{Local Invariants Do Not Generate Net Polarization  \label{sec:local_invariants} }

Constants of the motion that are local in space and time can never generate finite orbit width effects in real space, and, thus, they can never produce \emph{net polarization}.
One way of seeing this fact is that because the invariant is locally conserved, it is equal to it's own gyroaverage at every point in phase space, and, hence at every point along the particle trajectory.
In contrast, a nonlocal invariant is defined by an integral over the particle orbit with a integrand that varies from point to point.
These variations are responsible for finite orbit width effects.

Local invariants are generated by symmetries in space and time.
Hence, the only local invariants are: (1) the Hamiltonian, $\Ham$, for time-independent fields, (2) the linear momentum, $p_z$ for translation-invariant fields along coordinate $z$, and (3) the angular momentum, $\Pphi$, for rotation-invariant fields along angle $\phitor$.
The ``hairy ball'' theorem of topology states that any confinement region where the magnetic field is non-singular (and, hence, non-vanishing) must be topologically toroidal.
While the linear momentum, $p_z$, might be useful for studying simplified geometry, it is not conserved in toroidal geometry.
For toroidal confinement regions, this leaves only the Hamiltonian, $\Ham$, and the toroidal momentum, $\Pphi$, for angle $\phitor$.

For example, if the Hamiltonian is conserved due to time invariance, then
\begin{align}
\Ham&=\avgVel{\Ham}= \avgCyc{\Ham}
\\
 &=\mass\vcoord^2/2+\charge\potential=\mass\vpar^2/2+\Jcyc\Omgc + \charge\avgCyc{\potential}
\end{align}
at all points along the trajectory.
As shown in Sec.~\ref{sec:applications}, for the isotropic Maxwell-Boltzmann distribution,  which only depends on the Hamiltonian, the electric polarization and diamagnetic polarization must precisely cancel.
Net polarization is only generated by an additional dependence on the magnetic moment.
Hence, for the anisotropic Maxwell-Boltzmann distribution, there is a net polarization proportional to the  temperature anisotropy. 

For the same reason, a shifted Maxwellian distribution that depends on linear or angular momentum also cannot produce net polarization. 
For example, if angular momentum is conserved due to translation invariance in angle $\phitor$, then 
\begin{align}
\Pphi&=\avgVel{\Pphi}= \avgCyc{\Pphi}
\\
&=\mass\Vv\cdot\partial_\phitor \Vx=\avgCyc{\mass\Vv\cdot\partial_\phitor \Vx}
\end{align}
at every point in phase space, so, once again, there are no net finite orbit width effects.
However, in this case, the net polarization must also account for inertial forces, viscous forces, and friction forces.
If the flows are of first or higher order, then the electric and diamagnetic polarization will cancel to zeroth order, but this dominant balance will ultimately receive second or higher order corrections.
It is only if the flows are zeroth order and have substantial gradients that the electric and diamagnetic polarization do not cancel at zeroth order.

\subsection{Sources and Collisions \label{sec:kinetic_eqs} }
 
The kinetic equation is simply
\begin{align}
d\pdf/d\tcoord = \partial_\tcoord\pdf +\partial_{\zcoord^i} \left(\dot\zcoord^i \pdf \right)=\ColOp[\pdf]+\Source [\pdf]
\end{align}
where $\zcoord^i$ are arbitrary phase space coordinates, $\dot\zcoord^i$ is the time rate of change of the coordinates along the trajectory, $\ColOp[\pdf]$ is the collision operator, and $\Source[\pdf]$ is a source and/or sink of particles.
Both of these operators are functionals of the PDF and, are, in general, nonlinear.

Now, assume that one can transform to a set of adiabatic coordinates $\Zcoord=\{\Rcoord ,\Vpar, \Jcyc,\gyrophase\}$, with Jacobian $\Jac=\det{\partial \zcoord/\partial\Zcoord}$, where the gyrokinetic coordinates $\Zcoord^i(\tcoord,\zcoord)$ eliminate the fast motion of the gyrophase.
Now the kinetic equation for the PDF in these coordinates, $\PDF(\tcoord,\Zcoord)$, can be written as 
\begin{align}
d\PDF/d\tcoord &= \partial_\tcoord\PDF +\partial_{\Zcoord^i} \left(\dot\Zcoord^i \PDF \right)
+\partial_\gyrophase(\Omgc \PDF)\\
&= \ST^{-1} \ColOp[\ST\PDF]+  \ST^{-1} \Source .
\end{align}
If the evolution of the gyrophase, $\gyrophase$, is much more rapid than other physical processes, then, to lowest order, one can average the kinetic equation to find
\begin{align}
\avgCyc{d\PDF/d\tcoord} =\left[ \partial_\tcoord +\partial_{\Zcoord^i}  \dot\Zcoord^i  \right]\avgCyc{\PDF}=\avgCyc{ \ST^{-1}\ColOp[\ST\PDF]} + \avgCyc{\ST^{-1}\Source}.
\end{align}

Collisions are also important in determining the kinetic response because they add dissipation to the collisionless response.
Moreover, sources and collisions are both nonlinear processes (even if sometimes being approximated as linear), and this then impacts the solution for the PDF in various ways.
Thus, even when sources and collisions are weak, they can generate strong modifications to the PDF over time.

Once additional physics such as nonlinearities in amplitude, collisions, and sources are included, these effects begin to play a role in the result.
Because these effects are typically assumed to be weak, at the first order there is a clean separation of terms, but higher order corrections will mix these terms.
Nonetheless, the velocity moments of these corrections must be consistent with drift fluid theory.
Thus, additional corrections to the electric polarization and the diamagnetic polarization will be expressed through the drifts due to the viscous pressure tensor, the ram pressure, and the friction forces;  i.e. Eq.~\ref{eq:df-chargedensity-pol} in Sec.~\ref{sec:drift-fluid}.

Regardless of the form of the solution, if the drift kinetic expansion converges and is correctly applied to a given order $\epsilon^\porder$, then it will reproduce the correct solution to the same order.
The higher order terms will have the correct form and will be generated by perturbations of the lower order terms in the equation.

\subsection{Collisionless Equilibrium  \label{sec:collisionless} }

When considering the collisionless kinetic equation, which is determined by the advection operator alone, then there is a homogeneous solution, representing the degrees of freedom in the initial condition for the PDF.
In this case, the PDF can be any function of the slow coordinates $\Zcoord_0=\{\Rcoord,\Vpar,\Jcyc,\gyrophase_0:=\gyrophase-\Omgc\tcoord\}$.
There is no need for the form of the solution to have any relationship to a gyroaverage.

However, a complete solution of the kinetic equation requires consideration of sources and collisions in addition to advection.
If one integrates the sources in time from zero particles, i.e. $\PDF=0$, then the form of the solution will be completely determined by the kinetic response to the sources.
In the absence of collisions, the kinetic equation reduces to $d\pdf/d\tcoord=\Source$.
Integrating from zero particles at time zero, then the solution is simply
\begin{align}
 \pdf  = \int \Source d\tcoord = \int \Source d\ell /\Vcoord
\end{align}
which is the time average of the source along a trajectory from the entrance to the exit of the particle. In the second form, it is the average along the pathlength $\ell$ traveled at velocity $\Vcoord$.

\subsubsection{What form is natural for an equilibrium? }
The gyroaveraged part $\avgCyc{\PDF}$ is determined by the combination of the slower processes of collisions, advection, and sources.
Typically, collisions and advection by the parallel velocity are assumed to be zeroth order, while drifts and sources are assumed to be of first order in $\epsilon$. 
In the adiabatic coordinates, $\Zcoord=\{\Rcoord,\Vpar,\Jcoord,\gyrophase\}$, the gyroaveraged  kinetic equation becomes
\begin{align} 
 \avgCyc{ d \PDF/d\tcoord} = \avgCyc{\ST^{-1} \Cop[\ST \PDF]} + \avgCyc{\ST^{-1} \Source[\ST \PDF]}.
\end{align}
Note that,  even if an exact transformation to adiabatic coordinates cannot be found, the approximation
\begin{align} 
 \avgCyc{ d \PDF/d\tcoord} \simeq \avgCyc{ d/d\tcoord} \avgCyc{  \PDF} 
\end{align}
still holds to relatively high order due to the small size of the gyrovarying part $\PDFvary$.

The fact that each term is averaged implies that the lowest order solution typically has the form of a ratio of gyroaverages: the average source divided by the average escape rate
\begin{align}
\PDF = \avgCyc{ \ST^{-1} \Source} /\avgCyc{ \ST^{-1} \nu_{\eff}}.
\end{align}
The effective escape rate can have second and higher order corrections due to multiple factors in addition to gyroaveraging, such as higher corrections to the drifts, or to time dependent fluctuations.

For example, if there is a steady drift flow,  e.g. in the parallel or perpendicular direction, then the dominant solution will be determined by
\begin{align} 
\VV\cdot  \nabla_\Rcoord \PDF\simeq  \avgCyc{\ST^{-1} \Source}.
\end{align}
Since the lowest order drift $\VV$ can itself be expressed as a gyroaverage, e.g. $\VV_E=\avgCyc{\VE}\times\bhat/\Bfield$, then the result is a ratio of gyroaverages.
For a simple 1D variation where $\VV\cdot  \nabla_\Rcoord=\avgCyc{V^s}\partial_s$, the result can be written as 
\begin{align}
\PDF(s) - \PDF  (0) = \int_0^s  ds' \avgCyc{ \ST^{-1}\Source}/\avgCyc{\Vcoord^s}.
\end{align}
As another example, within quasilinear theory the effect of turbulence is to add an effective turbulent diffusion operator in the form
\begin{align} 
\nabla_\Rcoord \cdot \avgCyc{ D_{\rm eff} } \cdot  \nabla_\Rcoord \PDF\simeq  \avgCyc{\ST^{-1} \Source}.
\end{align}
For a simple 1D variation where $\nabla_\Rcoord \cdot \avgCyc{ D_{\rm eff} } \cdot  \nabla_\Rcoord =\partial_s   \avgCyc{ D_{\rm eff}} \partial_s$, the solution now takes the form
\begin{align}
\PDF(s) - \PDF  (0) =
\int_{0}^s ds' \avgCyc{D_{\rm eff}}^{-1} \int_{0}^{s'} ds'' \avgCyc{ \ST^{-1}\Source} 
.
\end{align}

\subsubsection{How large is the gyrovarying part of the PDF?}

For the gyrovarying part, $\PDFvary:=\PDF-\avgCyc{\PDF}$, collisions and drifts only make a small correction to the linear response, so the lowest order gyrovarying kinetic equation is simply
\begin{align}
\Omgc\partial_\gyrophase \PDFvary\simeq \Svary
.
\end{align}
For the purposes of the discussion here, the gyrovarying part of the source, $\Svary$, can be considered to include anything on the right hand side, such as the effects of collisions.
This can be solved by expressing any quantity in terms of  Fourier harmonics in gyrophase, e.g. for the PDF
\begin{align}
\pdf (\tcoord,\zcoord)=\ST \sum_\mindex\PDF_\mindex(\tcoord,\Zcoord) e^{i\mindex\gyrophase}
\end{align}
where the harmonics are given by
\begin{align}
\PDF_\mindex(\tcoord,\Zcoord) :=\oint \ST^{-1}  \pdf (\tcoord,\zcoord) e^{-i\mindex\gyrophase} d\gyrophase/2\pi = \avgCyc{\ST^{-1} \pdf e^{-i\mindex\gyrophase}}.
\end{align}
Hence, the kinetic equation becomes
\begin{align}
 \left[ \partial_\tcoord +\partial_{\Zcoord^i}  \dot\Zcoord^i +i\mindex\Omgc \right] \PDF_\mindex=\left(\ST^{-1}\ColOp[\ST\PDF]\right)_\mindex +  (\ST^{-1}\Source)_\mindex.
\end{align}
Neglecting collisions and drifts, then, to lowest order, the $\mindex\neq0$ harmonics are simply given by
\begin{align}
\PDF_\mindex=(\ST^{-1}\Source)_\mindex/i\mindex\Omgc.
\end{align}
Hence, this gyrophase-dependent correction is small and is of order $\sim \Source_\mindex/\mindex\Omgc\PDF_0<\CO(\epsilon_\perp\epsilon_\|/\mindex)$ for any given harmonic.
The key point is that the response to each harmonic must be treated separately.

To estimate the size of this correction, we must specify the size of the source.
First,  the gyro-varying terms are of the form $\gyro^\mindex\nabla^\mindex\Source$, so they are at least first order in $\epsilon$ or higher.
Second, the source injection rate, $\nu_S:= \Source/ \PDF  $ is assumed to be at most first order relative to the gyrofrequency, $\nu_S/\Omgc= \Source/\PDF\Omgc\sim \CO(\epsilon)$.
Hence, the gyrovarying terms of the PDF are at least second order in $\epsilon$ or higher.
In fact, the source injection rate is typically assumed to be one order higher than parallel advection, which makes it second order, $\nu_S/\Omgc\sim\CO(\epsilon^2)$.
This implies the gyrovarying terms of the PDF are at least third order in $\epsilon$ or higher.
Thus, for magnetized plasma applications, high enough accuracy can often be obtained by neglecting the higher harmonics generated by the source and simply setting $\Source\simeq \Source_0=\avgCyc{\Source}$. This is equivalent to assuming that the source itself is an exact gyroaverage.

\section{Conclusion \label{sec:conclusion}}

In this work,  the \emph{diamagnetic polarization paradox} was clearly defined and explained.
The apparent $2\times$ discrepancy between the diamagnetic polarization density in the real space (DK/DF) approach vs. the guiding center (GC), gyrokinetic (GK), and gyrofluid (GF) approaches can be explained by carefully accounting for all finite Larmor radius effects, including the specification of the \emph{zeroth order \reference distribution, $\pdf_0$}. 
The well-known GC/GK/GF result is that half of the contribution arises from transforming the guiding center density, $\Density(\Rcoord)$,  to  the real space density, $\density(\xcoord)$. 
In this work, it was shown that, in the real space picture, the other half of the contribution arises from the fact that the guiding center density, $\Density(\Rcoord)$, must be compared to the zeroth order density, $\densitystart(\xcoord)$, defined to be the fluid density in the limit of vanishing Larmor radius, $\kperp\gyro\rightarrow 0$. The combination of the two gyroaverages, required to perform the full mapping $\densitystart(\xcoord)\rightarrow \Density(\Rcoord)\rightarrow \density(\xcoord)$, yields the full diamagnetic polarization density.   

As explained in Sec. \ref{sec:collisionless}, a collisionless kinetic equilibrium must be the gyroaverage of the PDF in the limit of zero gyroradius, $\pdf_0$.
Moreover, as explained in Sec. \ref{sec:local_invariants}, when collisions are important, they require the equilibrium to be a function of local constants of the motion, such as the Hamiltonian, $\Ham$, and the toroidal momentum, $\Pphi$.
Any function of local constants of the motion is equal to its own gyroaverage at every point in phase space.
In this case, $\PDF:=\avgCyc{\pdf_0}=\pdf_0$, and, again, due to locality, $\pdf=\PDF=\pdf_0$.
Thus, local distributions must have zero net polarization.
For example, the isotropic Maxwell-Boltzmann distribution, which is only a function of the Hamiltonian, has no net polarization.
In contrast, the anisotropic Maxwell-Boltzmann distribution, which has an additional dependence on the magnetic moment, has a net polarization that depends on the temperature anisotropy.

In Part II, we will define and resolve  the \emph{Spitzer polarization paradox}, a direct corollary of the Spitzer magnetization paradox.
We will show that a careful analysis of symplectic guiding center theory completely resolves this paradox.

\appendix

\section{Guiding Center Coordinates \label{sec:gc_coordinates} }

The first step is to define  cylindrical velocity space coordinates $\Vv=\{\vpar,\vperp,\gyrophase\}$, where $\vpar$ is the velocity parallel to $\bhat$, $\vperp$ is the magnitude of the velocity perpendicular to $\bhat$, and $\gyrophase$ is the gyrophase. The gyro-orbits are ``diamagnetic'' so that a positively charged particle  traverses a left-handed circle, as shown in Fig.~\ref{fig:Larmor_orbit}.  Thus, the gyrophase, $\gyrophase$, is naturally defined as the angle in the plane perpendicular to the magnetic field that rotates in the left-handed sense. If one defines a right-handed coordinate system using orthonormal unit vectors $\{\ahat, \bhat, \chat\}$ that satisfy $\ahat\times  \bhat=\chat$, then, in a reference frame that is moving with the drift velocity, the gyrophase can be defined via  $\tan{(\gyrophase)} =- \Vv\cdot\ahat/\Vv\cdot \chat$. 

The guiding center coordinate transformation, $\CT_\GC: \zcoord=\{\Vx,\Vv\} \rightarrow \Zcoord= \{\VRcoord,\Vpar,\Jcyc,\gyrophase\}$, is defined by
\begin{align} \label{eq:gc_coordinates}
\Vx  &=\VR +\Vgyro &
\Vgyro &= \gyro  \cos(\gyrophase) \ahat +\gyro \sin(\gyrophase)\chat \\
  \Vv &=\VV + \Vw &
 \Vw&\simeq -w \sin(\gyrophase) \ahat +w \cos(\gyrophase) \chat 
  \label{eq:gc_velocity}
\end{align}
where $\VV$ is the drift velocity of the guiding center and 
\begin{align}
\Vw:=\left.d\Vgyro/d\tcoord\right|_\VR \simeq \Omgc \Vgyro\times \bhat.
\end{align}
 is the velocity around the guiding center. 
 Usually, the drift velocity $\VV$ is considered to be first order in $\kcoord\gyro$.
 Since $\Vpar\simeq \vpar$ and $\Vw\simeq\Vv_\perp  $ to lowest order, for certain expressions, the distinction between them can be neglected.

\section{Velocity Angle Average \label{sec:vspace-avg} }
Moments of the PDF, such as the density, are defined by an average over velocity space, and, hence, an average over the the direction of the perpendicular velocity.  
This velocity space phase angle average is also the lowest order approximation to the  gyroaverage.
While the velocity space average is not equal to the orbit average, it is an essential part of calculating any velocity space moment and the polarization density, in particular.

Velocity space moments such as the real space density, $\density$, and real space pressure, $\press$, average over  the direction of the perpendicular particle velocity, $\Vv=d\Vx/d\tcoord$, in real space, $\Vx$.
 This also yields the lowest order approximation to the orbit average.
 If the velocity space angle is defined  by the cylindrical angle of the perpendicular velocity via
 \begin{align}
 \gyrophase=\arctan{\left(-\Vw\cdot \ahat/\Vw\cdot\chat\right)},
 \end{align}
then the velocity angle average is defined by
\begin{align} \label{eq:gyroaverage_x}
\avgVel{\gfun }  = \oint \gfun (\xcoord,\vpar,\mu,\gyrophase)d\gyrophase/2\pi. 
\end{align}
The varying part is defined by 
\begin{align} \label{eq:gyrovary_R}
\varyVel{\gfun } =\gfun(\xcoord,\vcoord)-\avgVel{\gfun}.
\end{align}

\section{Gyroaverage \label{sec:gyroaverage} }

The gyroaverage is defined by the time average over the particle orbit and, thus, is simply an average over the gyrophase. 
Although there is only a single physical orbit average, there are several different approximations of the gyroaverage that depend on the coordinate system that is being used (defined in Appendix~\ref{sec:gc_coordinates} and Sec.~\ref{sec:gk_trans}) and on the order of approximation to the exact particle trajectory. 
A few useful approximations are discussed in the following.
 
 The lowest order approximation to the gyroaverage is simply the velocity angle average, discussed in Appendix \ref{sec:vspace-avg}.
A more accurate approximation to the gyroaverage is the average over the gyrophase  in guiding center coordinates  
\begin{align} \label{eq:gyroaverage_R}
\avgCyc{\Gfun }  = \oint \Gfun (\Rcoord,\Vpar,\Jcyc,\gyrophase)d\gyrophase/2\pi. 
\end{align}
For example, gyrofluid theory utilizes gyroaveraged  fluid moments of the distribution function in guiding center coordinates, such as the guiding center density, $\Density$, and pressure, $\Press$. In addition, the guiding centers respond to forces determined by the gyroaverage of the potentials in guiding center space, e.g. $\avgCyc{\phi}$. 
The gyrovarying part is defined by 
\begin{align} \label{eq:gyrovary_R}
\varyCyc{\Gfun }  =\Gfun(\Zcoord(\xcoord,\vcoord))-\avgCyc{\Gfun}  .
\end{align}

The zeroth order gyroaveraging operator will be defined by the integral
\begin{align} \label{eq:gyroaverage_Rbar}
\SJ_0[\Gfun] := \avgCyc{ \Gfun }  = \oint \Gfun (\Rcoord,\Vpar, \Jcyc,\gyrophase)d\gyrophase/2\pi
\end{align}
and not through the series for the Bessel function.  
More generally, we define the Fourier harmonics through the integral form
\begin{align} \label{eq:gyroaverage_Rbar}
\SJ_n[\Gfun] := \avgCyc{ e^{-in\gyrophase} \Gfun }  = \oint e^{-in\gyrophase} \Gfun (\Rcoord,\Vpar, \Jcyc,\gyrophase)d\gyrophase/2\pi
\end{align}
and not through the Bessel function series.

The notation is motivated by the fact that the gyroaveraging operation for the case of constant $\VB$.
In this case, $\Vpar=\vpar$ and it is simple to express for a function of real space coordinates, $\{\xcoord, \vpar, \mu\}$, that is otherwise independent of gyrophase, $\gyrophase$, e.g. through a dependence on $\Vvperp$.  
In this case, the gyroaverage of such a function is given by the expression
\begin{align}
\avgCyc{\ST^{-1}\gfun(\xcoord,\vpar,\mu) }  =\SJ_0\circ\ST^{-1} \gfun=   J_0(b)\ST^{-1}\gfun 
\end{align}
where  $J_0(b)$ is the  Bessel function of order 0 with the operator $b=-i\gyro\nabla_\perp$ as its argument; in Fourier space, the argument is simply $b= \kperp\gyro$.
Similarly, the real space gyroaverage of a function of $\Rcoord$ alone is
 \begin{align}
\avgVel{\ST \Gfun(\Rcoord,\Vpar,\Jcyc) }  = \SJ_0\circ \ST  \Gfun= J_0(b)\ST \Gfun 
 .
\end{align}

The  previous definitions of the gyroaverage use a low-order approximation to the actual trajectory.  
A higher-order approximation to the trajectory can  be obtained by performing the gyroaverage in the higher-order adiabatic coordinate system, $\GKcoord$, e.g. described in Sec.~\ref{sec:gk_trans}.
This  gyroaverage  will be denoted via
\begin{align} \label{eq:gyroaverage_Rbar}
 \avgCycBar{ \Gfun }  = \oint \bar \Gfun (\Rbar,\bar \Vpar,\bar \Jcyc,\bar \gyrophase)d\bar \gyrophase/2\pi. 
\end{align}
The gyrovarying part is defined by 
\begin{align} \label{eq:gyrovary_R}
 \varyCycBar{\Gfun } =\bar\Gfun(\GKcoord(\xcoord,\vcoord))- \avgCycBar{\Gfun}.
\end{align}
However, when working to linear order in amplitude, this more accurate calculation is not required. 

This more accurate gyroaverage can rarely be evaluated in closed form because it involves a nonlinear coordinate transformation. It does have a simple expression to first order in the amplitude of the electric potential
\begin{align} 
 \avgCycBar{\Gfun }  &= \avgCyc{ \Gfun  }  + \varyCyc{\charge \phipert} \partial_{\Jcyc}\avgCyc{ \Gfun  }   /\Omgc+\dots
\\
&\simeq \avgCyc{ \Gfun} +  \charge \phipert  \partial_{\Jcyc} \avgCyc{ \Gfun} /\Omgc - \avgCyc{\charge \phipert   }  \partial_{\Jcyc}  \avgCyc{ \Gfun} /\Omgc .
\label{eq:gyroavg_first}
\end{align}
This results in a simple expression to 2nd order in $\kperp\gyro$
\begin{align} \label{eq:gyroavg_FLR}
 \avgCycBar{\Gfun } \simeq \avgCyc{ \Gfun  }+ \tfrac{1}{2} \charge \gyro^2 \nabla_\perp ^2\phipert \partial_{\Jcyc}\avgCyc{ \Gfun  }  /\Omgc .
\end{align}

\section{Gyrokinetic Coordinate Transformation \label{sec:gk-transformation} }

For constant $\VB$, the lowest order coordinate system in velocity space has identical $\gyrophase$, $\Vpar=\vpar$, and $\Jcyc\Omgc=\mu\Bfield=\mass\vperp^2/2$.
Since the Jacobian, $\CJ=\Omgc$, is constant, scalars and probability densities transform in the same manner.

The pushforward transformation, $\ST^{-1}:\zcoord\rightarrow\Zcoord$,  is given by
\begin{align}
\Gfun(\Rcoord,\Vpar,\Jcyc,\gyrophase)&=\ST^{-1}\gfun  \\
&= \int  \delta^3(\Vx-\VR-\Vgyro(\Zcoord))\gfun(\xcoord,\Vpar,\Jcyc,\gyrophase)d^3\xcoord \\
&=\int e^{i\Vk\cdot(\VR +\Vgyro)}\gfun(\kcoord,\Vpar,\Jcyc,\gyrophase) \frac{d^3\kcoord}{(2\pi)^{3/2}} \\
&=\int   e^{i\Vk\cdot \VR  +i\kperp\gyro\sin{(\gyrophase-\gyrophase_\kcoord)}}\gfun(\kcoord,\Vpar,\Jcyc,\gyrophase) \frac{d^3\kcoord}{(2\pi)^{3/2}} 
\\
 &=  \sum_\mindex \int e^{i\Vk\cdot\VR+ i\mindex (\gyrophase-\gyrophase_\kcoord) } J_\mindex(b)  \gfun(\kcoord, \Vpar, \Jcyc,\gyrophase)  \frac{d^3\kcoord}{(2\pi)^{3/2}} 
\end{align}
where $b=\kperp\gyro$.
Here, the angle   $\tan{(\gyrophase_\kcoord)}:=-\Vk\cdot\ahat/\Vk\cdot\chat$ is defined through the definition
(compare to the definition of $\Vw$ in Eq.~\ref{eq:gc_coordinates})
\begin{align}
\Vkperp:= -\kperp \sin(\gyrophase_\kcoord) \ahat +\kperp \cos(\gyrophase_\kcoord) \chat.
\end{align}

If $\gfun_0(\xcoord,\vpar,\Jcyc)$ is independent of gyrophase, then the Fourier harmonics in gyrophase take a simple form 
\begin{align}
\Gfun_\mindex  (\Rcoord,\Vpar,\Jcyc) &= e^{- i \mindex \gyrophase_\kcoord } \SJ_\mindex \gfun_0:= \oint e^{- i\mindex \gyrophase }\ST^{-1} \gfun_0 d\gyrophase/2\pi\\
&=
 \int    e^{- i \mindex \gyrophase_\kcoord }   J_\mindex (b) \gfun_0(\kcoord,\Vpar,\Jcyc)   \frac{d^3\kcoord}{(2\pi)^{3/2}} 
 \end{align}
that we use to define the operators $\SJ_\mindex \gfun$ and $e^{-i\mindex\gyrophase_\kcoord} $ in Fourier space.
This then yields the result
\begin{align}
\Gfun(\Zcoord) &= \sum_\mindex   e^{   i\mindex \gyrophase}\Gfun_\mindex=\sum_\mindex   e^{   i\mindex  (\gyrophase - \gyrophase_\kcoord  ) }\SJ_\mindex  \gfun_0 .
\end{align}
 For arbitrary dependence on gyrophase, the result is
\begin{align}
\Gfun(\Zcoord) &= \sum_\mindex   e^{   i\mindex \gyrophase}\Gfun_\mindex=\sum_\mindex   \sum_n e^{   i\mindex (\gyrophase-\gyrophase_\kcoord)}\SJ_{ \mindex-\nindex}    \Gfun_{\nindex} .
\end{align}

The pullback transformation, $\ST:\Gfun(\Zcoord)\rightarrow\gfun(\zcoord)$,   is given by
\begin{align}
\gfun(\zcoord)&=\ST  \Gfun (\Zcoord)=\Gfun(\Zcoord(\zcoord)) \\
&= \int  \delta^3(\Vx-\VR-\Vgyro(\zcoord))\Gfun(\Zcoord,\vpar,\Jcyc,\gyrophase)d^3\Rcoord \\
&=\int e^{i\Vk\cdot(\Vx -\Vgyro)}\Gfun(\kcoord,\vpar,\Jcyc,\gyrophase) \frac{d^3\kcoord}{(2\pi)^{3/2}} \\
&=\int   e^{i\Vk\cdot \Vx  -i\kperp\gyro\sin{(\gyrophase-\gyrophase_\kcoord)}}\Gfun(\kcoord,\vpar,\Jcyc,\gyrophase) \frac{d^3\kcoord}{(2\pi)^{3/2}} 
\\
 &=  \sum_\mindex \int J_\mindex(b)  \Gfun(\kcoord, \vpar, \Jcyc,\gyrophase) e^{i\Vk\cdot\Vx - i\mindex (\gyrophase-\gyrophase_\kcoord) } \frac{d^3\kcoord}{(2\pi)^{3/2}} 
\end{align}
where, again, $\tan{(\gyrophase_\kcoord)}:=-\Vk\cdot\ahat/\Vk\cdot\chat$.

If $\Gfun_0(\Rcoord,\Vpar,\Jcyc)$ is independent of gyrophase, then the Fourier harmonics of $\gfun$  are  
\begin{align}
\gfun_\mindex  (\Rcoord,\Vpar,\Jcyc) &= \SJ_{-\mindex}\Gfun_0= \oint e^{  i\mindex \gyrophase }\ST^{-1} \Gfun_0 d\gyrophase/2\pi
\\
&=
\oint e^{ - i \mindex ( \gyrophase- \gyrophase_\kcoord) }   J_{-\mindex}(\kperp\gyro) \Gfun_0(\kcoord,\Vpar,\Jcyc)    \frac{d^3\kcoord}{(2\pi)^{3/2}} 
 \end{align}
which defines the operators $\SJ_{-m}\Gfun$ and $e^{ i\mindex\gyrophase_\kcoord} $ in Fourier space. This yields the simple result
\begin{align}
\gfun(\zcoord) &= \sum_\mindex   e^{   i\mindex \gyrophase}\gfun_\mindex=\sum_\mindex   e^{   i\mindex (\gyrophase-\gyrophase_\kcoord)}\SJ_{-\mindex}    \Gfun_0 .
\end{align}
For arbitrary dependence on gyrophase, the result is
\begin{align}
\gfun(\zcoord) &= \sum_\mindex   e^{   i\mindex  \gyrophase   }\gfun_\mindex=\sum_{\mindex,\nindex} e^{   i\mindex (\gyrophase-\gyrophase_\kcoord)}\SJ_{\nindex-\mindex}    \Gfun_{-\nindex} .
\end{align} 
 
\vspace{12pt}
\begin{acknowledgements}
The author would like to thank C. Bolton for inspiring research on the diamagnetic polarization ``factor of 1/2" paradox and for many interesting discussions on this topic.  The author would also like to thank A. J. Brizard,  B. I. Cohen, A. M. Dimits, M. A. Dorf, and G. W. Hammett for a number of illuminating discussions and helpful commentary on this manuscript.  In particular, G. W. Hammett stressed the connection to the Spitzer paradox.

This work, LLNL-JRNL-862399, was performed under the auspices of the US Department of Energy (DOE) by LLNL under Contract DE-AC52-07NA27344 and was supported by the DOE Office of Fusion Energy Sciences.
\end{acknowledgements}  

\section*{Data Availability}
The data that support the findings of this study are available from the corresponding author
upon reasonable request.


\bibliography{Polarization_GK-DF_v1.bib}	

\vfil
\eject
\end{document}